# *In Planta* Tattoo and Kirigami Sensors for Self-Powered Monitoring of Vapor Pressure Deficit and Growth Dynamics

*Nafize Ishtiaque Hossain [a,b], Kundan Saha [a,b], Atul Sharma [a,b], Sameer Sonkusale [a,b]\**

[a] Sonkusale Research Labs, Tufts University, Medford, MA 02155, USA.

[b] Department of Electrical and Computer Engineering, Tufts University, Medford, MA 02155, USA.

\*Corresponding author email: sameer.sonkusale@tufts.edu

**Abstract:**

We report a scalable, self-powered *in planta* sensor platform for continuous monitoring of plant hydration and growth. The system integrates two components: a leaf-mounted tattoo sensor for estimating vapor pressure deficit (VPD) and a kirigami-inspired strain sensor for tracking radial stem growth. Uniquely, the tattoo sensor serves a dual function—measuring temperature and humidity beneath the leaf surface while simultaneously harvesting power from ambient moisture via a vanadium pentoxide ($V_2O_5$) nanosheet membrane. This moist-electric generator (MEG) configuration enables energy-autonomous operation, delivering a power density of 0.1114 µW/cm². The $V_2O_5$-based sensor exhibits high sensitivity to humidity (4.2 mV/%RH) and temperature (1.02%/°C), enabling accurate VPD estimation for over 10 days until leaf senescence. The eutectogel-based kirigami strain sensor, wrapped around the stem, offers a gauge factor of 1.5 and immunity to unrelated mechanical disturbances, allowing continuous growth tracking for more than 20 days. Both sensors are fabricated via cleanroom-free, roll-to-roll compatible methods, underscoring their potential for large-scale agricultural deployment to monitor abiotic stress and improve crop management.

# INTRODUCTION:

Climate change poses a significant global challenge to agriculture worldwide. It affects crop yields, disrupts the food supply, and decreases nutritional quality of the produce grown, thereby making precision farming necessary [1]. Precision agriculture is a farming approach that leverages data and technology to optimize the use of water, fertilizer, and pesticides under different growing and environmental conditions to improve efficiency and crop productivity [2–4]. Recent advances in precision farming have focused heavily on remote sensing, satellite imaging, and drone-based technologies [5]. Direct monitoring methods that employ measurement of water and nutrients in soil and plants provides more fine-grained local information [6–11]. These are achieved using portable sensors that measure key parameters such as temperature, pH, humidity, facilitating the study of plant response to different environmental conditions [12–16]. By providing high-resolution, real-time measurements of these parameters, these technologies have the potential to predict plant stress and assist growers in taking corrective action before substantial losses occur [17,18]. Of all sensor options available for monitoring plant stress, skin-like [19] or tattoo [20,21] sensors have attracted significant research interest due to their non-invasive and flexible form factor, and their ability to capture multimodal plant health data [22–24]. These plant-wearable sensors are usually realized on PDMS [25], PET [26], and polyimide [27,28] substrates and have demonstrated impressive performance. However, the development of an integrated, self-powered multimodal tattoo sensor that can simultaneously monitor stressors and growth-related factors has not been demonstrated. Moreover, existing solutions rely on external power sources such as batteries that increase cost and worsen the environmental impact. Our objective is to develop a fully self-powered environmentally friendly wearable sensor suite for accurate monitoring of plant health and stress.

Prior studies suggest that plant transpiration rate is a reliable proxy for plant water stress [29,30] and significantly affected by vapor-pressure deficit (VPD) [31]. VPD incorporates temperature and relative humidity at the leaf surface and the surrounding air and indicates whether plants are exposed to dry and wet growth conditions. A high VPD signifies a dry state for the plant [32] and can also serve as a marker for biotic and abiotic stresses [33]. Additionally, plant stress can hinder radial stem growth, and chronic stress may even cause stem shrinkage [34]. Thus, monitoring the radial growth of the plant stem, especially in real-time, is important and to help understand how local plant stress is impacting its growth, allowing farmers to take proportional corrective action.

In this work, we propose an *in-planta* health monitoring system consisting of a leaf-tattoo-based sensor utilizing $V_2O_5$ nanosheet-based moisture electric generator (MEG) for real-time monitoring of relative humidity (RH). The $V_2O_5$ nanosheet based MEG also demonstrates temperature dependent voltage output thereby facilitating temperature sensing beneath the plant leaves. Both RH and temperature measurements enables calculation of the vapor pressure deficit (VPD) in plants. Secondly, a eutectogel kirigami-based wrap-around strain sensor utilizing gel-like material is used for monitoring plant's radial growth. The resistive response of the eutectogel kirigami strain sensor yields real-time measurements of stem diameter. The entire system consisting of leaf tattoo and wrap-around strain sensor is connected to a custom low-power electronic unit powered by the MEG itself. The electronic module concurrently processes signals from the eutectogel kirigami strain sensor and the MEG tattoo, enabling a comprehensive assessment of plant's physiological state. To evaluate the sensor performance across physiologically relevant conditions, real plant measurements were obtained under controlled imposition of abiotic stressors namely water deficit and salinity treatments. This innovative self-powered *in planta* sensor suite is conceptually depicted in Figure 1. We envision that the integration of the self-powered plant tattoo with a kirigami strain sensor and custom-designed low-power electronics for real-time plant monitoring will pave the way for more efficient plant surveillance and accelerate decision-making in both crop breeding and precision agriculture.

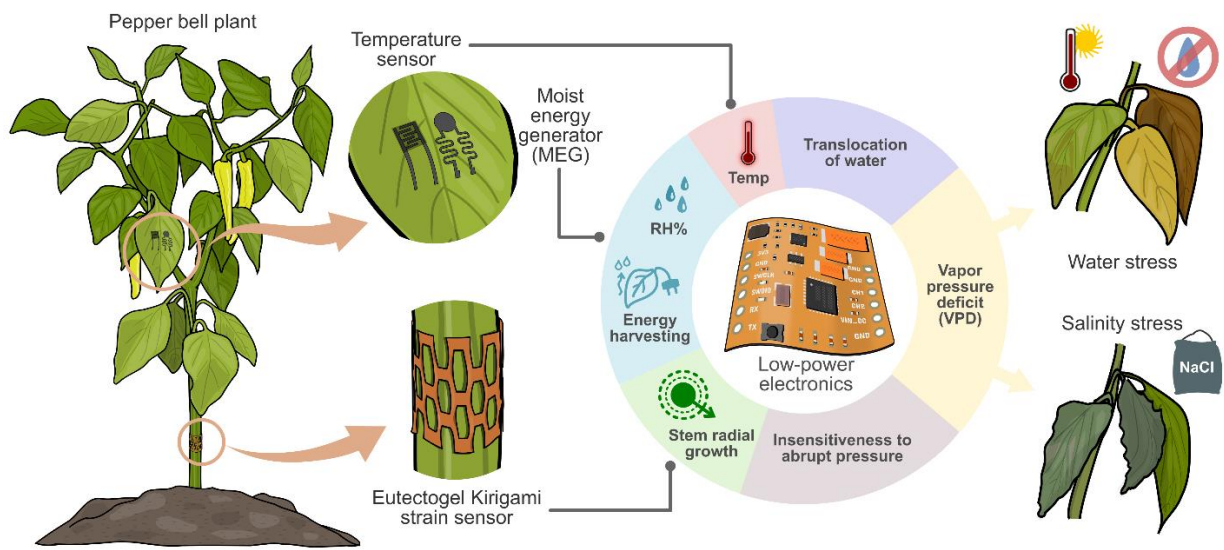

**Figure 1.** Self-powered *in planta* sensor for suite for vapor pressure deficit (VPD) and radial growth measurement for assessment of water stress and salinity stress monitoring.

**RESULTS AND DISCUSSION:**

The detailed preparation procedure for the $V_2O_5$ nanosheet membrane for realization of MEG, Relative Humidity (%RH) and temperature sensing, and eutectogel for strain sensing, is described in the experimental section and briefly illustrated in Figures S1 and S2, respectively.

**Tattoo-based Leaf VPD Sensor and Eutectogel Kirigami Strain Sensor Fabrication:**

$V_2O_5$ nanosheet based MEG based leaf tattoo serves both to harvest energy from moisture and provide a measurement of relative humidity on leaf surface, both parameters are important for measurement of VPD. Temperature sensing is achieved by monitoring the temperature coefficient of resistance of $V_2O_5$ nanosheets across interdigitated electrodes. To fabricate the *in-planta* tattoo for the leaf and the stem, the tattoo's stencil was first created using transfer tape (Chartpak, DAF8), which was patterned in the Universal Laser System (ULS) according to a design file in AutoCAD Fusion 360 (.DXF format). The final tattoo design for the leaf and the stem, including size, is detailed in Figure S3. A serpentine pattern was used to minimize stress on the tattoo due to the plant growth that can cause external strain. The patterned stencil was then transferred onto the tattoo substrate paper. Graphene ink (ethanol as the solvent) was applied to form the interdigitated electrode for temperature sensing and a section of the MEG. This coating was cured at 50 °C for 5 hours. All electrodes were electrochemically cleaned to remove contaminants from the electrode surface, as described in our previous work [35,36]. In parallel, a PET sheet patterned for the second half of the MEG was coated with graphene ink on both sides and cured at 100°C for 30 minutes. The customized $V_2O_5$ nanosheet (dia.- 5mm) was then placed between the graphene-coated PET layer and the graphene-coated substrate layer in a sandwich format to complete the MEG assembly. A 10 μL aliquot of $V_2O_5$ dispersion was drop-cast onto the interdigitated temperature sensor layer and cured at 50°C for an additional 30 minutes. The tattoo carrier sheet was subsequently attached to the tattoo sensor and allowed to adhere overnight before application to the plant leaf or stem surface. The detailed fabrication procedure for plant-tattoo fabrication is outlined in Figure 2a.

The stem sensor is designed to monitor increasing strain from radial growth using a gel of a eutectic solvent with high ionic conductivity (namely eutectogel) on a kirigami-patterned polyimide substrate. Similar to the tattoo, the kirigami pattern was designed in AutoCAD software to be created on a polyimide (PI) sheet (Cole-Parmer, 125.0 μm thickness) under a ULS laser machine. Laser cutting introduced a thin layer of carbonization on the cutting edge of the kirigami-patterned PI sheet, which, if not removed, could cause irregularity in strain sensor behavior. Therefore, the kirigami-patterned PI substrate was thoroughly washed with isopropyl alcohol

(99.9%) to remove laser-induced carbon flakes along its cut edges. The patterned PI sheet was then rinsed with ethanol (95%), followed by DI water, and dried at 50°C. The two edges of the kirigami pattern were coated with Ag/AgCl paste (Applied Ink Material, AG-500) and cured at 100°C for 40 minutes.

Strain sensing is achieved by monitoring the impedance of the eutectogel, which is a nonvolatile ionically conductive electrolyte within a stable gel scaffold. Eutectogels enable long-term usage without any need for additional encapsulation, such as for ambient protection [37]. Unlike hydrogels, it remains stable for extended periods in ambient conditions and allows air and vapor permeability and preserving breathability of the host plant [38]. These deep eutectic solvents (DES) comprise choline chloride (ChCl) and ethylene glycol (EG), incorporated into a gelatin scaffold. Importantly, EG promotes fewer but larger and more dynamic helices, while choline chloride (ChCl), due to its kosmotropic nature, aids in forming flexible chain bundles, enhancing the gel's stretchability. These features make eutectogels excellent functional materials for wearable plant strain sensors [39]. The preparation procedure for the eutectogel is detailed in the experimental section based on our earlier reported work [40]. 100 μL of the viscous gel was drop-cast onto the air plasma-etched kirigami substrate, followed by spin-coating at 500 rpm for 1 minute to achieve a uniform gel coating. The eutectogel-coated kirigami structure was then stored in a dry environment (relative humidity less than 5.0%) overnight before being attached to the plant stem. The detailed fabrication procedure of the eutectogel kirigami strain sensor is illustrated in Figure 2b.

The Finite Element Method (FEM) simulation of the plant tattoo sensor and the kirigami strain sensor using the ANSYS software has been shown in figure 1c and supporting information Figure S4 respectively. The detailed experimental setup and discussion are provided in the Supporting information Note S1. The usage of the serpentine structure helps the tattoo sensor to be insensitive to strain both from high degree of bending and stretching making them a practical solution in real environmental applications, also confirmed by the FEM simulation results. Similarly, the FEM simulations shows that the kirigami structure of the strain sensor helps to distribute the strain on the structure evenly – strain is divided uniformly among the distributed kirigami paths. This configuration makes the strain sensor anisotropic such that abrupt strain activity vertical to the axial strain, dissipates the resulting strain slowly along different kirigami paths – kirigami also provides less surface area for any vertical strain impact compared to bulk eutectogel based implementation.

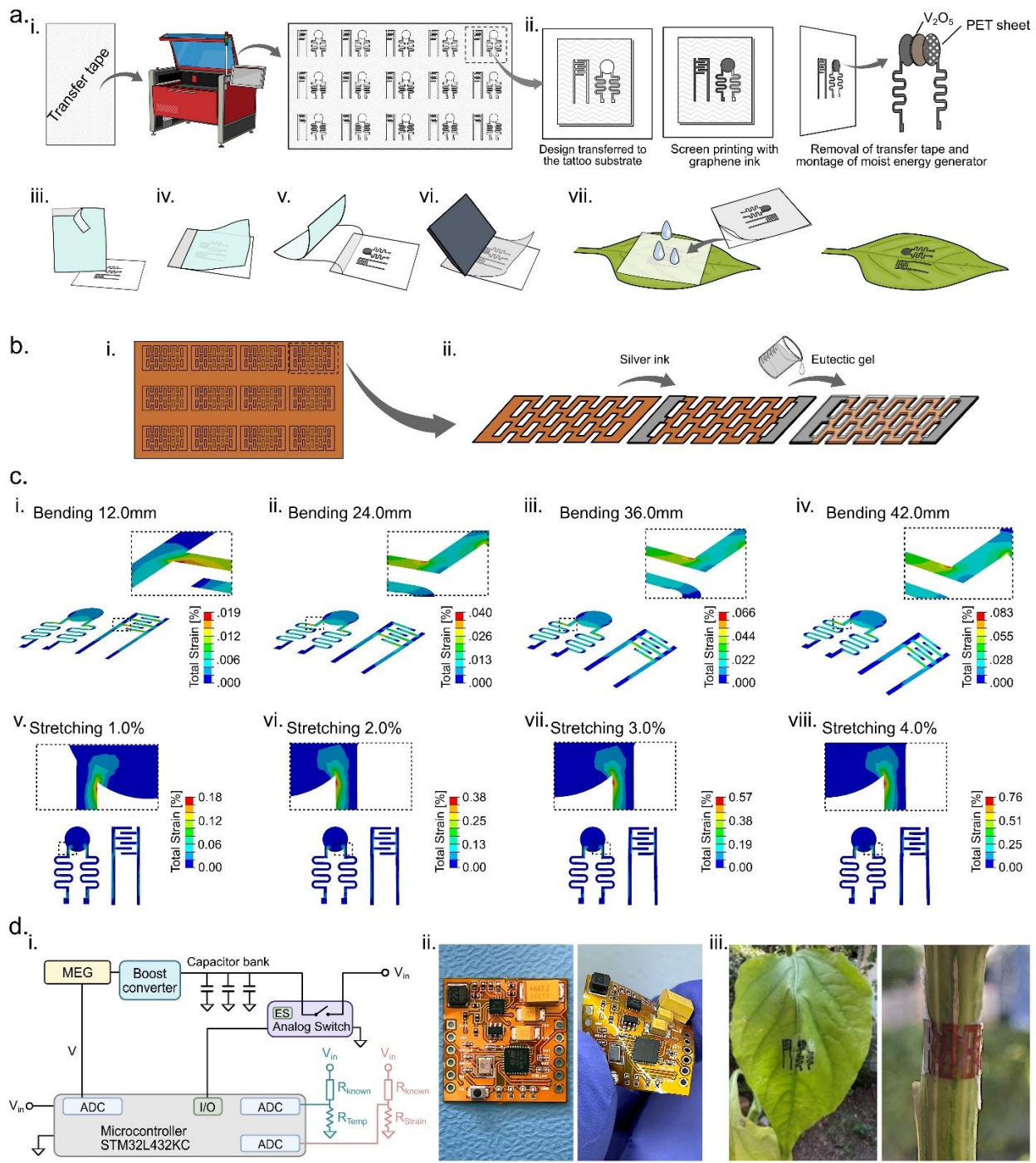

**Figure 2.** (a) Leaf tattoo making process. (i) Transfer tape insertion in laser machine and pattern transfer on a roll basis; (ii) stencil separation from the main roll, attachment of the stencil to the tattoo paper, graphene ink attachment, fabrication of MEG by insertion of V$_2$O$_5$ sheet and graphene-coated PET sheet and drop casting of V$_2$O$_5$ in interdigitated humidity sensor.; (iii)

the opening of the tattoo carrier sheet is peeled off; (iv) the carrier sheet initial part is attached to the tattoo paper top part; (v) the protector part of the tattoo transfer tape is removed carefully; (vi) using squeegee the carrier paper of the tattoo is attached to the main tattoo paper; (vii) the tattoo paper is attached to the plant leaf and the using the water soaked cloth the tattoo paper is pressed for 10 minutes on top of the leaf, and then the tattoo is transferred to the leaf; (b) preparation of eutectogel kirigami Strain Sensor; (i) the kirigami pattern created on polyimide sheet using laser induction; (ii) Ag/AgCl paste is coated on the two kirigami edge, and (iii) the eutectogel solution drop cast and spin-coated on top of the kirigami structure and rested overnight. (c) Mechanical deformation analysis using ANYSYS software of the plant tattoo sensor. (d) Implementation of the self-powered *in planta* sensor and systems; (i) the functional block diagram of the low-power electronic circuit; (ii) developed low-power circuit; (iii) *in planta* tattoo temperature and MEG-based humidity Sensor Suite; (iv) kirigami strain sensor wrapped around the plant stem.

It is noteworthy to mention that in the case of the temperature sensor, there may be some dependency due to strain from the use of straight interdigitated finger electrodes (Figure 2c(i-iv)). This could have been avoided by adopting a serpentine design for electrodes as well. However, the serpentine design introduces another level of variability in the geometry, particularly in the separation of the interdigitated electrodes, which would then affect the impedance (both resistance and capacitance) of the interdigitated sensor [41,42]. This would significantly affect the temperature measurement compared to using a straight inter-digitated electrode design for the temperature sensor. As a compromise, we used large area straight interdigitated fingers for temperature measurement from $V_2O_5$ nanosheets assembled onto them.

The functional block diagram of the developed low power flexible circuit is shown in Figure 2d(i), and the detailed PCB design trace is shown in Figure S5. The custom-made electronics are depicted in Figure 2d(ii) and the full fabrication details is described in experimental section. The real time implementation of the plant tattoo and the kirigami strain sensor on the real bell pepper plant is shown in figure 2d(iii).

**Moist Electric Generator (MEG) Characterization and Measurement of Energy Harvesting and Humidity Measurement:**

The MEG relies on the the use of stacked $V_2O_5$ nanosheet membranes synthesized from bulk exfoilated bulk $V_2O_5$ crystal. The orthorhombic crystal structure of pristine $V_2O_5$ exhibits a chained structure that is interconnected by oxygen bridges, ultimately forming square-based pyramids near the vanadium atoms. Specifically, the vanadyl ($O_v$), bridge ($O_b$), and chain ($O_c$) structures contribute to electronic conductivity and ion transport. In this paper, pristine $V_2O_5$ was exfoliated into ultra-thin nanosheets by treating with hydrogen peroxide, forming a homogeneous dispersion, as described in our recent work [43]. The SEM image in Figure 3a shows that the flakes have dimensions of 100-200 nm. Zeta potential analysis of the $V_2O_5$ dispersion (11.25 µg/mL) yielded a value of -15.70 mV (Figure 3b), indicating the presence of negatively charged surfaces on the nanosheets. The $V_2O_5$ nanosheet dispersion was vacuum-filtered onto a PTFE support membrane. Upon drying, a free-standing lamellar membrane was formed that detached from the support membrane. A photo of the free-standing membrane is shown in Figure 3c. A cross-sectional scanning electron microscope (SEM) image (Figure 3d) reveals a uniform stacking pattern of the nanosheets. The X-ray diffraction (XRD) pattern in Figure 3e also displays a distinct peak at d = 1.4 nm, corresponding to the interlayer spacing along the c-axis (001 plane) [16,17]. This peak underscores the highly ordered stacking of nanosheets within the $V_2O_5$ nanosheet membrane, affirming the integrity of the targeted structural arrangement. For the fabrication of MEG, a seived PET sheet is created using a laser cutter, and both sides of the PET sheet are coated with graphene ink and placed on top of the $V_2O_5$ nanosheet. When water evaporates off the leaf surface, they come in contact with the $V_2O_5$ nanosheests *via* the graphene-coated sieved PET sheet (Figure 3f). The summary of the working mechanism of the $V_2O_5$ MEG is shown in Figure 3g. The water molecules then dissociate into $H_3O^+$ and $OH^-$ inside the nanochannels of the $V_2O_5$, primarily due to the negative surface charge of the $V_2O_5$ nanosheets [44,45]. As the nanochannel height is smaller than the Debye length (961 nm) [46], electrical double layer overlaps, causing the $H_3O^+$ cations to sweep through the cation selective nanochannels. This unipolar movement of hydronium ions, creates a potential difference, generating an open circuit potential proportional to the water content, and which ultimately drives an electric current through an external load connected across it. This forms the basic principle of operation of the proposed MEG.

The developed MEG has been tested under various humidity levels, ranging from 30% and 90 %RH. Exposure to different humidity levels results in varying generated voltage and current, as shown in Figures 3h and 3i. In the humidity chamber, the humidity was controlled by a combination of a humidifier and a dehumidifier, as discussed in our earlier work [7,8]. Calibration experiments were performed using a Keithley SourceMeter 2450. We also investigated the effect of varying thicknesses of the $V_2O_5$ membrane, revealing that as the thickness of the $V_2O_5$ membrane increases, both voltage and current increase until saturation is reached. From Figures 3j and 3k, we conclude that a 145 nm thick $V_2O_5$ sheet, within our current design, provides the highest power output, making it suitable for energy harvesting in our tattoo MEG.

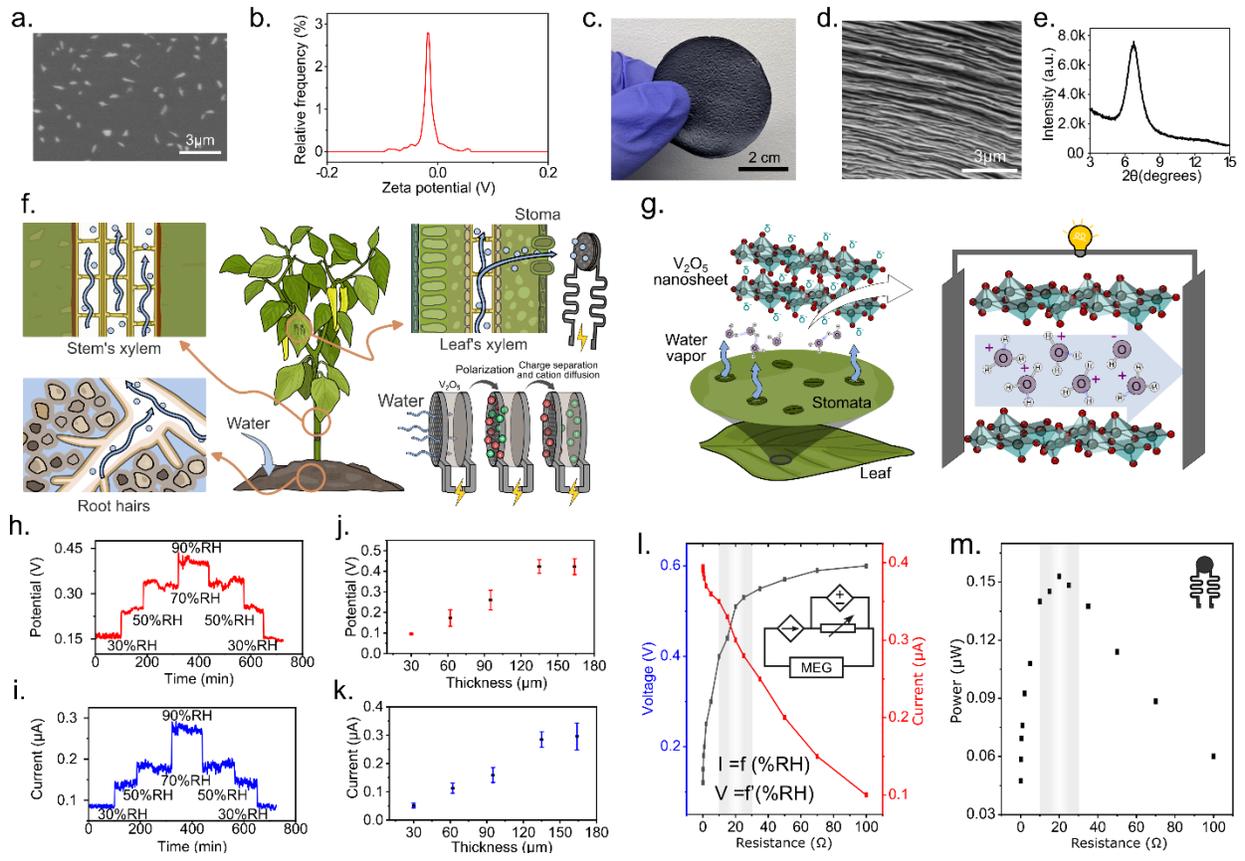

**Figure 3.** (a) The SEM image of the $V_2O_5$ nanosheets; (b) Zeta potential distribution of the $V_2O_5$ nanosheets; (c) photo of the $V_2O_5$ membrane ; (d) cross-section of the $V_2O_5$ membrane ; (e) XRD of the $V_2O_5$ membrane; (f) Working principle of the tattoo based MEG and humidity sensor; (g) energy harvesting mechanism of $V_2O_5$ nanosheets; the voltage response (h) and (i) current response of the MEG for various humidity level; voltage (j) and current; (k) variation based on the thickness of the $V_2O_5$ membrane, (l) current and voltage generation performance under different resistance loading; (m) maximum power point tracking (MPPT) of the MEG.

Figure S6 shows the open circuit voltage (OCV), the short circuit current, and the I-V characteristics of the developed $V_2O_5$-based MEG for 30% RH. Energy efficiency of the MEG is measured as the percent fraction of the input energy converted to output energy.

To calculate the output energy, the following procedure is adopted from previously published reports [47]. Initially, MEG is placed in a sealed bottle containing a fixed amount of DI water and a $V_2O_5$-incorporated membrane, and kept at 25.0 °C and 60% relative humidity for 24 hours without disturbance. During these 24.0 hours, the open-circuit voltage and short-circuit current are calculated to obtain the output energy, which is 0.0589 J. After 24.0 hours, the weight of the bottle is again measured, and the amount of evaporated DI water is determined, which is 0.415 g. By converting the evaporated water mass to moles and applying the heat of evaporation (44 kJ/mol), the input energy was calculated as 1.0136 kJ.

The MEG generates micro-watt of power which is not sufficient to power the readout electronics module. To address this issue, we used a Maximum Power Point Tracking (MPPT) based approach driving the boost converter. Circuit theory dictates that maximum power transfer occurs when the source impedance matches the load impedance—in our case, the input impedance of the BQ25570 boost converter. To find the MEG's internal impedance, we measured voltage and current across various fixed resistances at 90 % RH (Fig. 3l), then plotted power versus resistance curve (Fig. 3m). The peak power occurred with a 20 Ω load, so we designed the boost converter's input network to match that impedance. The boost converter then charges a supercapacitor bank with the harvested energy, which in turn drives the electronics module for readout from the leaf-humidity sensor that is based on the MEG's terminal voltage itself. When the analog switch connects the supercapacitor bank to the circuit, the microcontroller exits deep sleep mode and subsequently it enters low-power mode. The first action of the microcontroller in low power mode is to measure the terminal voltage of the MEG. Based on a predefined calibration curve established during programming (Figure S7), the microcontroller detects the relative humidity beneath the leaf surface and in the open air. A voltage divider-based resistance measurement technique has been adopted for low-power applications, particularly for temperature and strain sensing [48].

**Temperature Sensor and Eutectogel Kirigami Strain Sensor Characterization:**

The temperature sensor made from $V_2O_5$ functional layer across the working area of the interdigitated electrodes operates across a temperature range from 15°C to 60°C. The sensor demonstrates both repeatability and reproducibility, as shown in Figure 4a. As previously discussed, $V_2O_5$ possesses an orthorhombic nanosheet structure and exhibits n-type semiconducting behavior with a bandgap of ~2.2 eV [49]. This relatively small bandgap facilitates thermally activated polaron hopping between $V^{4+}$ and $V^{5+}$ sites, driven by increased lattice vibrations at elevated temperatures. When the temperature increases, the resistance decreases, and the opposite happens when a low temperature is applied, indicating that $V_2O_5$ exhibits a negative temperature coefficient (Figure 4b). The temperature sensor also displays a linear behavior, which is consistent with the existing literature [50]. Although the temperature sensor shows hysteresis of 0.05% during abrupt temperature changes, this does not impact our measurements as such rapid environmental fluctuations are rarely encountered in practical scenarios [51,52]. The temperature sensor optimization based on the $V_2O_5$ concentration and volume are illustrated in Figure S8.

The leaf-to-tattoo impedance data were collected using the methodology illustrated in Fig. 4c, which was adapted from a previously reported approach [53]. Briefly, three sets of tattoos—designated as A, B, and C—were affixed to the leaf. The leaf-to-tattoo impedance was measured using tattoo B. A sine wave current ($I_{AC}$ = 10–20 µA) generated by a function generator was applied to tattoo A at varying frequencies ranging from 1 Hz to 1 kHz, while tattoos B and C were maintained under open-load conditions. Here, the AC voltage ($V_{AC}$) is measured between tattoos B and C, where the observed voltage drop arises solely from the impedance between the tattoo B interface and the underlying leaf tissue, comprising both xylem and phloem. Specifically, the impedance $Z_{e2}$ is calculated as the ratio of the measured AC voltage ($V_{AC}$) to the applied AC current ($I_{AC}$), i.e., $Z_{e2} = \frac{V_{AC}}{I_{AC}}$. Figure 4d illustrates the impedance profile of the leaf–tattoo interface, which exhibits a decreasing trend with increasing frequency. Additionally, the reduced impedance at the leaf–tattoo interface facilitates real-time energy harvesting and reliable data collection.

The strain sensor on the stem utilizes a kirigami design, which is known for its ability to sustain more significant stress than bulk materials [54]. Our previous research has demonstrated that deep eutectic solution (DES)-based gelatin gels are effective strain sensors [39]. In this study, we applied a thin layer of eutectogel on the kirigami surface, enhancing its stress-bearing capability, as illustrated in Figure 4e, up to a strain level of approximately 250%. The stress-strain curve of the eutectogel-integrated kirigami structure is better than previously reported bulk eutectogel based strain sensors [40]. When strain is applied to the eutectogel kirigami strain sensor, the resistance changes, and the corresponding relative resistance change versus strain curve is shown in Figure 4f. From this curve, we extracted the gauge factor defined as ration of normalized resistance to the applied strain $\left(\frac{\frac{\Delta R}{R_0}}{\varepsilon}\right)$. The kirigami-based strain sensor exhibited a gauge factor of 1.5 at lower strain levels, decreasing to 0.6 at higher strains. This improvement, compared to our previous gauge factor of 0.6 [39], might be attributed to adapting the kirigami structure and using resistance as the measurement parameter rather than capacitance.

To assess the performance of the strain sensor to measure plant stem growth, the eutectogel kirigami strain sensor was wrapped around 3D-printed cylindrical structures of different diameters. A Form 3B resin printer was used for printing using a Biomed Clear resin (ver. 2.0). The degree of curvature was calculated using the formula: $r = \frac{360S}{2\pi\theta}$, where θ is the angle of curvature, r is the radius of curvature, and S is the arc length. The length of our eutectogel kirigami strain sensor is 2.1 cm, representing the arc length S. The sensor's response to various bending strains is presented in Figure 4g. Additionally, Figure S2 shows the orientation of the eutectogel component forming a network. Note that the bending strain was calculated using a different formula, $\varepsilon = \frac{t}{2r_b}$, where *t* is the total thickness of the sensor (substrate plus functional layer, totaling 205 μm, with a standard deviation of 0.8 μm), and $r_b$ is the bending radius [55,56]. As bending strain increases, the relative resistance also rises. From the resistance versus bending strain response, we can determine the angle of curvature to which the strain sensor is attached, allowing us to calculate the radius of the cylindrical surface (i.e., the plant stem).

For measurement and real-time applications, we wrapped the eutectogel kirigami strain sensor around a plant stem, measuring relative resistance to derive the angle of curvature and, ultimately, the change in stem diameter. The sensor was also tested under various temperature and humidity conditions, as shown in Figures 4h and Figure 4i respectively. To mitigate variability in strain sensor readings due to environmental factors, we applied a passivation layer of Kapton tape (Uline, 10 μm thickness) over the kirigami strain sensor (Figure 4j). This tape serves a dual purpose: blocking UV rays [57,58] from sunlight that may cause altering the chemical properties/performance of the eutectogel layer, and reducing artifacts caused by temperature and humidity fluctuations.

We assessed the strain sensor's response time using a high-precision 32-bit sigma-delta analog-to-digital converter (ADS1362, Texas Instruments). As shown in Figure 4k, the sensor achieves a stable resistance response within approximately 7 (±0.23) ms when subjected to bending strain. For practical applications, a 12-bit successive approximation (SAR) ADC from the

STM32L432KC microcontroller was used due to its lower power consumption, to extract strain sensor data using a simple voltage divider circuit.

We also investigated the impact of external disturbances on the performance of the strain sensor. Two strain sensors were placed on the plant stem, each with an eutectogel functional layer—one featuring a kirigami structure and the other a flat structure of the same dimensions. We performed an experiment in a controlled environment, where a fly is allowed to land on the strain sensor – this is expected to cause a rise in the resistance. For the experiment, flies are collected using a honey trap. Then, the plant with the eutectogel kirigami strain sensor and the bulk eutectogel strain sensor is applied on the plant, which has been placed inside a confined chamber with the electrical connections so that the real-time resistance can be achieved. The stem beneath the two sensors (in the same vicinity) are coated with honey before placing the sensors. The flies are released in the chamber, and when they sit on the top of two of the strain sensors, the disturbance is recorded as a change in resistance. The kirigami-based sensor exhibited a slower and lower magnitude response to this disturbance than the flat (non-kirigami) substrate, as shown in Figure 4l. Importantly, any such disturbances that caused a shift in the baseline, was further automatically corrected by the microcontroller unit. The reduced and delayed resistance response of the kirigami structure to disturbances is attributed to the anisotropic nature of the strain sensor. Reduced landing surface area and multiple load distribution paths inherent in its design, reduces the ability of the fly to impact strain measurement, as discussed earlier.

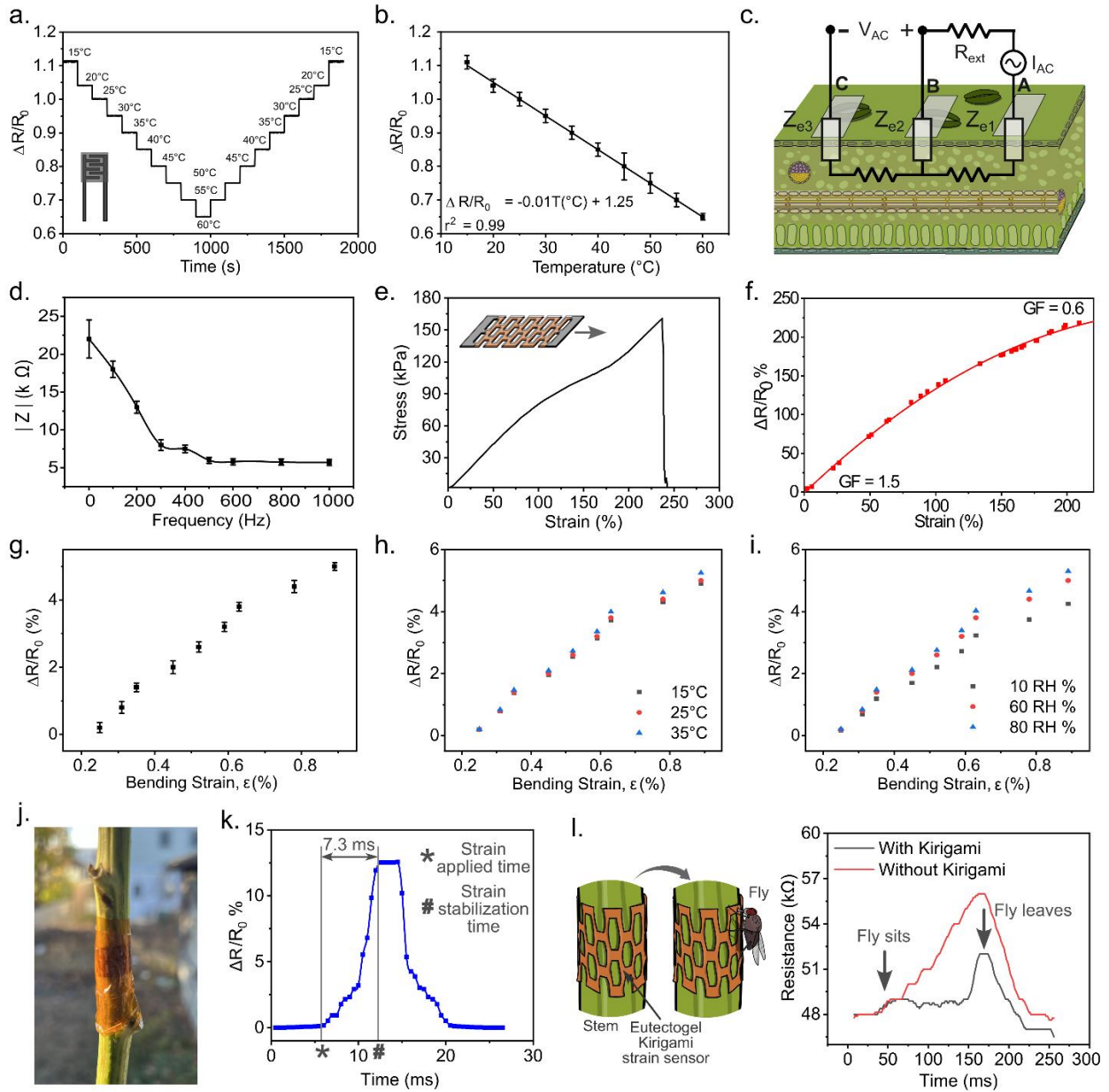

**Figure 4.** (a) Dynamic resistance change of the temperature sensor with respect to time; (b) calibration curve of the temperature sensor; Characterization of the eutectogel kirigami strain sensor, (c) The leaf to tattoo impedance measurement scheme; (d) Impedance profile of the leaf to tattoo interface; inset: possible equivalent circuit for the leaf-to tattoo interface; (e) Stress-strain response of the strain sensor; (f) Relative resistance versus strain response of the eutectogel strain sensor; (g) Relative resistance vs. bending strain response; (h) Relative

resistance vs. bending strain response for varying temperature with respect to constant 60 RH%; (i) Relative resistance vs. bending strain response for varying humidity with respect to constant 25°C temperature; (j) Kapton tape-covered kirigami strain sensor during real time application; (k) Relative resistance response when instantaneous strain is applied; and (l) resistance variation of the kirigami based and plane strain sensor when a fly sits on the strain sensor during attachment to the plant stem.

**Water Translocation Analysis:**

Water translocation is a significant parameter to monitor water uptake and plants response to stress. Conventionally, it has been monitored using lysimeter [59], leaf water potential [60], and isotopic tracers-based methods [61], however they are often invasive, costly, low-throughput, or indirect, limiting their ability to provide continuous, high-resolution, and non-invasive insights into plant water dynamics [62–65]. To measure water translocation in plants, we utilized two tattoo sensors: one attached to the lowest leaf and the other to a leaf near the top, spaced 17 cm apart vertically. The experiment was conducted on a healthy bell pepper plant, with watering at midday (12:30 PM). We observed that the relative humidity (RH) level in the lower leaf increased earlier than in the upper leaf. It took approximately 225 minutes for the RH levels of both leaves to equalize (Figure 5a). Additionally, we employed a self-powered readout circuit (indicated by the dotted line). In this case, a self-powered array of MEGs in tattoo form (3x3, total of 9) was attached to the leaf and connected to a boost converter circuit. By positioning the self-powered tattoos at two distinct locations on the plant, the developed system enables monitoring of water translocation from the roots to the upper leaves which is an important plant physiology.

**VPD Analysis:**

To measure vapor pressure deficit (VPD), the sensor is placed on the mid-section of the leaf, with the sensing element positioned beneath it on a full-grown bell pepper plant. (For demonstration purposes, the tattoo sensor is shown on the top of the leaf in Figure 2d(iii).) All VPD measurements are conducted with the sensor placed at the mid-leaf position. VPD represents the difference between the vapor pressure of saturated air beneath the leaf surface and that of the surrounding atmosphere. The vapor pressures for both saturated and open air can be calculated using the following formula [66].

$$VPsat = 0.6107 \times 10^{\frac{7.5 T_1}{237.3 + T_1}} \quad (1)$$

$$VPair = 0.6107 \times 10^{\frac{7.5 T_a}{237.3 + T_a}} \times \frac{RH}{100} \quad (2)$$

In the above equations, $T_1$ represents the temperature beneath the leaf surface, while $T_a$ denotes the open-air temperature. Tests were conducted at the Tufts University campus (42.408222 °N and 71.116402 °W) from September 20 to October 4, 2024. All reported values were calculated using the temperature and humidity sensor integrated into the tattoo, with tests performed in an open environment to validate the sensor's applicability in real-world conditions. The sensors were attached to 20 healthy plants, 20 water-stressed plants, and 20 salinity-stressed plants. Prior to the experiment, all 60 plants were acclimatized in the same environment, and the experiment commenced only when the vapor pressure deficit (VPD) levels of all plants were nearly uniform, with a standard deviation of 0.078 kPa observed. For the control (healthy) plants, the VPD exhibited oscillatory behavior; in contrast, the water-stressed plants showed a consistent increasing trend (Figures 5b and 5d), corroborating previous reports [7,67]. The rising VPD trend indicates rapid transpiration in the plants, while a declining VPD trend suggests saturation of the leaf surface [8]. High VPD values also indicate that the stomata of the plant leaves are closed, reflecting the plant's attempt to resist transpiration. Real-time observations of stomatal opening and closing are illustrated in Figure S9, captured using an optical microscope (Keyence, VH-7100).

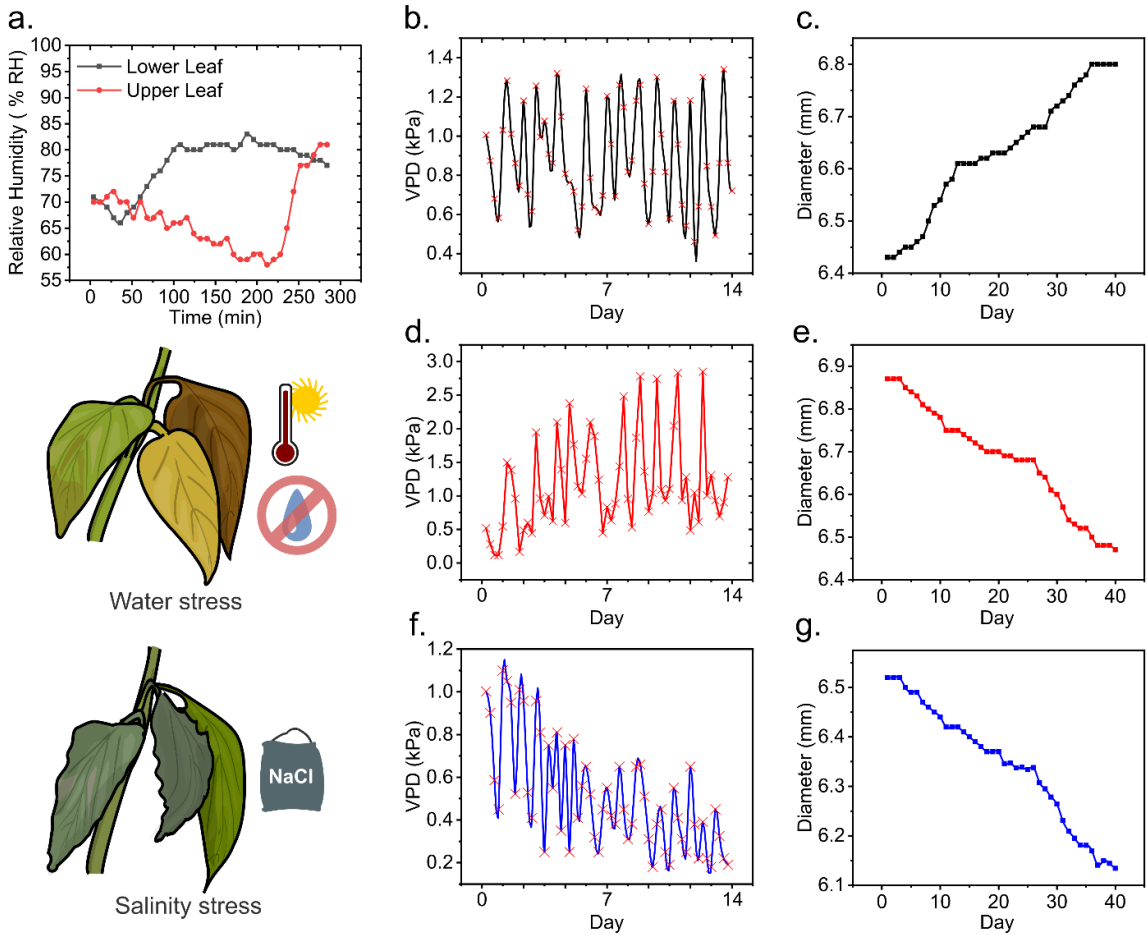

**Figure 5.** Real-time data collection from the *in-planta* sensor. (a) Translocation of water from the root to the leaf as the lower leaf obtains water early compared to the top leaf; (b) healthy plant VPD data; (c) healthy plant stem growth; (d) water-stressed plant VPD data; (e) water-stressed plant stem diameter; (f) salinity stress plant VPD data; (g) salinity stress plant diameter.

In the case of VPD, the red cross marks represent the VPD data collected using a self-powered circuit, and the continuous lines represent the VPD data collected from external powered device. With the harvested energy, we collected five vapor pressure deficit (VPD) data points over 24

hours—three during the day and two at night. To validate the accuracy of the data, we attached two tattoo sensors to the same leaf; one set was self-powered, while the other relied on an external power source. The continuous lines in Figure 5b, 5d, and 5f represent the externally powered data (black for healthy plants, red for water-stressed plants, and blue for salinity-stressed plants), while the discrete red cross marks indicate the self-powered data. We also measured VPD in 20 pepper plants under salinity stress, applying 100 mL of 1M NaCl solution (each day). Unlike the water-stressed plants, those subjected to salinity stress exhibited a decrease in VPD compared to control plants (Figure 5f). This reduction might be attributed to decreased stomatal conductance and changes in humidity due to salinity presence [68]. As plants experience salinity stress, their ability to uptake water is compromised by osmotic stress, leading to stomatal closure and a subsequent decline in VPD [69]. Interestingly, during our experiments measuring relative humidity and temperature, we observed that the temperature beneath the leaf was consistently lower than that of the surrounding air. In contrast, the humidity below the leaf was higher (Figure S10). It is important to note that the humidity data obtained from the MEG were temperature-corrected using methods described in earlier reports [7,8].

**Stem Diameter Analysis:**

When water stress is applied to the plant, a consistent decrease in stem radial growth is observed, as shown in Figure 5e, compared to the healthy plant depicted in Figure 5c. A similar response is noted in the radial growth of plants subjected to salinity stress, illustrated in Figure 5g. The plant diameter reduction occurs due to physiological and structural changes, such as osmotic stress, ion toxicity, and most importantly, phytohormonal imbalance inside the plant due to the higher salt level. The resultant effect of these phenomena inhibits the cell expansion, leading to a halt in growth and, in extreme cases, a reduction in stem diameter [70,71]. These findings are consistent with previous reports [68,72].

**Robustness Analysis of Sensor Suite:**

The flexibility of the tattoo sensor was assessed by bending it at 10° and stretching it by 4% using motorized transition base (Thorlabs MTS50A-Z8) coupled with brushed motor controller (Kinesis KDC 101) at 2mm/s speed. The bending and stretching have been done for 100 cycles. Although such extreme bending and stretching are unlikely in real-world applications, these conditions were used to evaluate the calibration performance. The results indicated that the coefficient of variance (CV) for both the temperature and humidity sensors was less than 2.1% (Figures S11 and S12) which coincides with the results obtained from the FEM based mechanical deformation analysis. The tattoo sensor's humidity and temperature data have been validated using a commercially available DHT11 temperature and humidity sensor (while interfaced with Arduino Nano), while the data stem diameter value obtained from the kirigami strain sensor is validated using slide calipers (Mitutoyo). The data obtained from commercial sensors and instrumentation, and the data obtained from sensors and systems have a CV less than 0.6%. VPD measurements were taken from both a bent and an unbent tattoo sensor attached to the same leaf, utilizing an external power source for validation; the CV was found to be less than 1.0%. This low CV suggests that the sensor maintains performance even after significant bending and stretching.

The sensors (tattoo and kirigami based strain sensor) attached to the plant was exposed to natural light rain for 15 minutes, however no significant performance drop has been observed for energy harvester (/humidity sensor), temperature sensor and strain sensor, because of the protective carrier sheet layer in tattoo and Kapton tap encapsulation in eutectogel kirigami strain sensor. Both the tattoo sensor (temperature and humidity) and kirigami-based strain sensor have been tested for short-term (1 hour) and long-term (12 hours) continuous stability/durability tests, and the CV is less than 1.5% (Figure S13). For practical durability testing, the tattoo sensor was affixed to a plant leaf for 43 days until the leaf dried, making data extraction impossible (Figure S14). Figure S14 also indicates that the tattoo sensor did not adversely affect leaf growth and adapted to the leaf's expansion. Specifically, after installation, the leaf area increased by 2.71 cm², as measured using image processing techniques in MATLAB.

Additionally, a kirigami-based strain sensor was subjected to an extreme level of nearly 100% strain by stretching at a rate of 1 mm/s for 2.0 seconds, followed by a return to its original state after 1 second of rest (totaling 5.0 seconds). This cyclic stretching was repeated for 100 cycles (Figure S15), with only minimal hysteresis (less than 2%) observed. After completing these cycles, three sets of the eutectogel kirigami strain sensors were placed on plant stems (one control plant, one water-stressed plant, and one salinity-stressed plant). The sensors exhibited a response similar to that of pristine sensors for over 10 days. Detailed data on the stem diameters for these three sensors are provided in Table S1. The total sensor suite (one set of tattoo sensors and one set of kirigami-based strain sensor) costs less than $3.5 (and less than $20, including low-power readout electronics). However, the preparation of the sensor suite is time-consuming since it requires filtration of $V_2O_5$ nanosheets and requires an inert environment – this can be considerably improved using industrial scale methods of vacuum filtration and roll to roll printing.

**CONCLUSION AND FUTURE WORKS**

We present a leaf-conformal, self-powered hybrid tattoo capable of monitoring plant leaf temperature, humidity, water translocation, and vapor pressure deficit. We also demonstrated a eutectogel-based wrap-around kirigami strain sensor for monitoring the radial growth of plant stems. By establishing a scalable fabrication process that employs $V_2O_5$ as the sole functional material in the plant tattoo and integrates a eutectogel-based kirigami strain sensor, we successfully enabled self-powered detection of two critical plant biomarkers—vapor pressure deficit (VPD) and stem radial growth—without relying on any external power source. Moreover, our sensing platform simultaneously delivers stretchability, and conformability, with no discernible adverse effects on plant health. Leveraging these attributes, we integrated the plant tattoo with custom-designed, flexible, low-power electronics on the leaf surface, enabling long-term, real-time monitoring of plant growth and vapor pressure deficit (VPD). The *in-planta* monitoring results captured distinct day–night VPD rhythms along with corresponding variations in leaf surface temperature, highlighting the capability of our plant tattoo to track physiological signals relevant for real-world plant monitoring applications. Notably, the ability to transduce real-time plant physiological signals is particularly valuable for accelerating the identification of desirable phenotypes in the precision breeding of new plant varieties. Furthermore, our plant tattoo sensors can effectively distinguish and detect two major forms of abiotic stress: water stress and salinity stress. The platform also holds promise for powering other sensors—such as macronutrient sensors, phytohormone sensors, and biopotential sensors. The integration of multiple sensors with self-powered functionality offers farmers, agronomists, and plant biologists a valuable tool for gaining deeper insights into plant physiology under different growing conditions. In summary, we have provided a preliminary demonstration of a flexible, self-powered sensor system capable of monitoring plant conditions—such as humidity and temperature—as well as radial stem growth. In future, one could expand the platform towards breathable, self-powered, multimodal sensing tattoo capable of detecting plant nutrients, phytohormones, volatile organic compounds (VOCs), bioimpedance, and biopotentials. Moreover, one could expand the time frame of monitoring under diverse stress conditions and even advanced machine learning approaches to enable early identification of specific stress types.

## EXPERIMENTAL SECTION

### Materials and Methods

Vanadium pentoxide, hydrogen peroxide, graphene ink in ethanol, choline chloride, ethylene glycol, and a PTFE membrane are purchased from Sigma-Aldrich (USA). These materials are all of analytical grade and are used without further purification or modification.

### Preparation of $V_2O_5$ Membrane:

A 25.0 mL uniform dispersion of vanadium pentoxide ($V_2O_5$) was prepared by solubilizing 177.2 mg/mL of $V_2O_5$ in distilled water in a 500 mL beaker. This mixture was then placed in a cold-water bath to control the heat generated in the subsequent reaction step. To initiate the exfoliation process, 25.0 mL of hydrogen peroxide (30%) was slowly added to the $V_2O_5$ dispersion, resulting in an exothermic reaction. Successful reaction completion was indicated by the formation of a $V_2O_5$ foam, which was then diluted with 150.0 mL of deionized (DI) water. This mixture was then sonicated at 85 W continuously for 48 hours until a uniform suspension was achieved. A 12.0 mL aliquot of this mixture was vacuum filtered through a 0.20 μm PTFE membrane for 48 hours, yielding a uniform $V_2O_5$ membrane with a thickness of 145 μm. Finally, the freestanding $V_2O_5$ nanosheet membrane was cut to fit the dimensions required for the energy-harvesting working electrode. Membranes of various thicknesses were fabricated by adjusting the volume of the solution (Supporting Information, Figure S1).

### Preparation of Eutectogel:

Briefly, a molar ratio of choline chloride: ethylene glycol: water = 1:2:1 was prepared with continuous magnetic stirring and heating at 100°C under an inert $N_2$ environment. The stirring speed was maintained at 250 rpm to prevent the introduction of air bubbles, which could contribute to unwanted resistance variation in the sensor and affect reproducibility. Next, 22 wt % of gelatin was added to the deep eutectic solution to form a gel, with heating and stirring maintained. After 10 hours of continuous stirring and heating, it is prepared for drop-casting.

**Readout Electronics Design and Development:**

The main processing unit of the data acquisition system was the low-power microcontroller STM32L432KC. The microcontroller can stay alive in deep sleep mode and operate at ultra-low power levels. Through source code (which is written in Cube IDE), the power consumption of the microcontroller is kept up to 5μA (at 3.3V). This measurement was performed using the Nordic Power Profiler Kit II. The current and voltage generated from the $V_2O_5$-based-MEG are very low, making it impossible to operate the microcontroller with this voltage and current. For this reason, a boost converter was used in the readout circuit. The boost circuit charged two capacitors (660 μF and 220 μF), connected in parallel, with a total equivalent capacitance of 880 μF. The main element of the boost conversion is BQ25504, and it charges the 880μF at 3.3V. Once the two capacitors are fully charged and the voltage reaches 3.3V, an analog switch MAX4644 is closed. When this analog switch is closed, the fully charged equivalent 880μF capacitor supplies energy to the microcontroller through $V_{in}$.

Two low-power voltage buffer amplifiers were developed for the analog reading of the voltage generated by the MEG generator, effectively rejecting the noisy voltage signal produced by the MEG, which was ultimately used to extract humidity data. A voltage divider-based resistance measurement was adopted for temperature measurement to facilitate power management, as both

Wheatstone bridge-based resistance sensing and instrumentation amplifier-based resistance measurement consumed more power. For open-air temperature measurement, the microcontroller's internal temperature sensor was utilized. A separate MEG was built and kept in open air for open-air humidity measurement; this MEG was combined with the leaf MEG at the electronics board for energy harvesting.

It should be noted that using the internal RC oscillator of the microcontroller at low frequency enabled the readout circuit to achieve this performance despite the engineering trade-off of having a high system latency of 935 (±2) ms. In our design, we chose to allow this high system latency because the focus was on tracking plant parameters (VPD and radial growth) through the harvested energy rather than acquiring high-frequency and high-resolution data. The developed board included provisions for an external crystal oscillator for use in battery-powered applications. The real-time plant tattoo sensor and the eutectogel kirigami strain sensor affixed to the plant leaf and stem, respectively, are shown in Figure 2d(iii). The connection between the leaf tattoo and the kirigami strain sensor to the low-power electronic circuit was made using flexible copper wire. Just after the installation of the sensor and systems, the circuit is instantly primed with a charge capacitor initially.

**ASSOCIATED CONTENT**

The data supporting this article have been included in the Supplementary Information.

Detailed synthesis of $V_2O_5$ nanosheet and DES gel, dimensions of sensors, FEM analysis of kirigami strain sensor, design of PCB and component description, calibration curve of humidity and optimization of the temperature sensor, real implementation of sensor, microscopic characterization, sensor performance (bending and stretching analysis), and real-time performance and comparison of sensor.

## CONFLICTS OF INTEREST

The author declares no conflict of interest.

## AUTHOR CONTRIBUTIONS

**Nafize Ishtiaque Hossain:** Conceptualization, Methodology, Experimentation, Investigation, Data curation, Validation, Formal analysis, Writing-original draft writing, and editing. **Kundan Saha**: Experimentation, Investigation, Data curation, Validation, Formal analysis, Writing review, and Editing. **Atul Sharma:** Experimentation, Investigation, Data curation, Formal analysis, Writing-original writing, draft review, and editing. **Sameer Sonkusale:** Conceptualization, Methodology, Resources, Supervision, Project Administration, Funding Acquisition, Writing-Reviewing, and Editing. All authors have given their approval for the final version of the manuscript.

## ACKNOWLEDGMENTS

This research received partial support from the National Science Foundation (NSF) funded project "Large Area Distributed Real-Time Soil (DiRTS) Monitoring" under Grant 1935555. More recently, the authors acknowledge partial financial support of the Defence Advance Research Project Agency (DARPA) eXVi, and Tufts University's Graduate Student Research Competition Award 2024.

Supplementary Information

# *In Planta* Tattoo and Kirigami Sensors for Self-Powered Monitoring of Vapor Pressure Deficit and Growth Dynamics

*Nafize Ishtiaque Hossain [a,b], Kundan Saha [a,b], Atul Sharma [a,b], Sameer Sonkusale [a,b]\**

[a] Sonkusale Research Labs, Tufts University, Medford, MA 02155, USA.

[b] Department of Electrical and Computer Engineering, Tufts University, Medford, MA 02155, USA.

\*Corresponding author email: sameer.sonkusale@tufts.edu

# Table of Contents

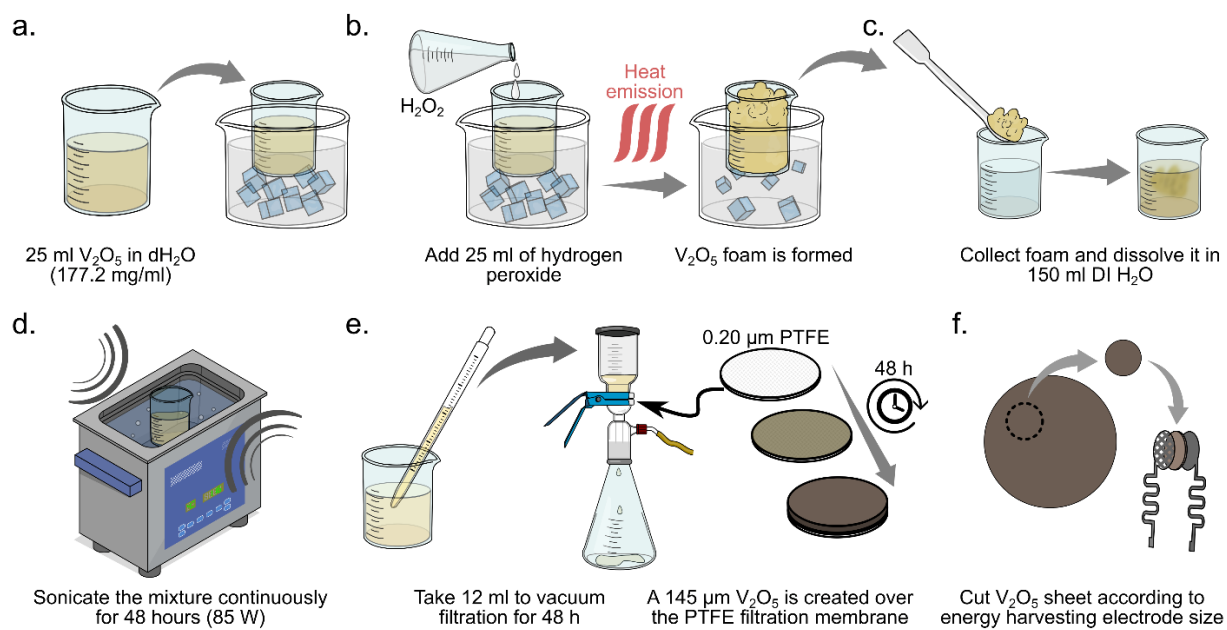

**Figure S1:** A step-by-step preparation of vanadium pentoxide ($V_2O_5$) membrane

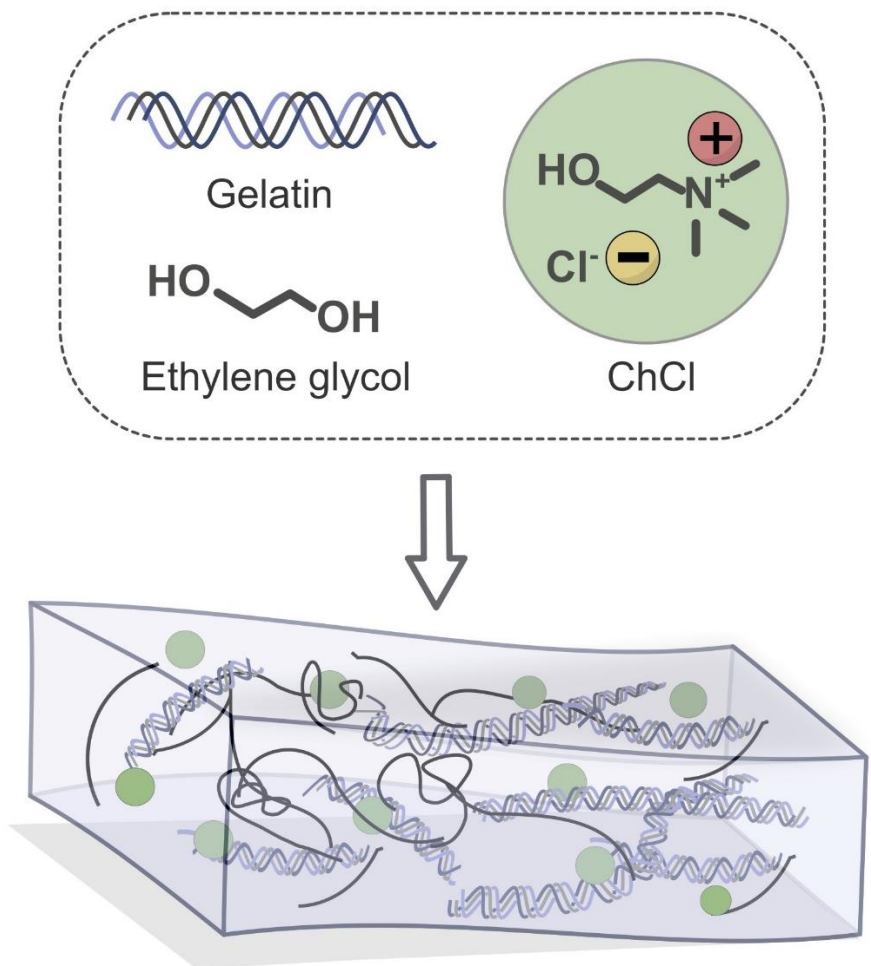

Figure S2: Preparation and Structural Representation of DES Gel Network

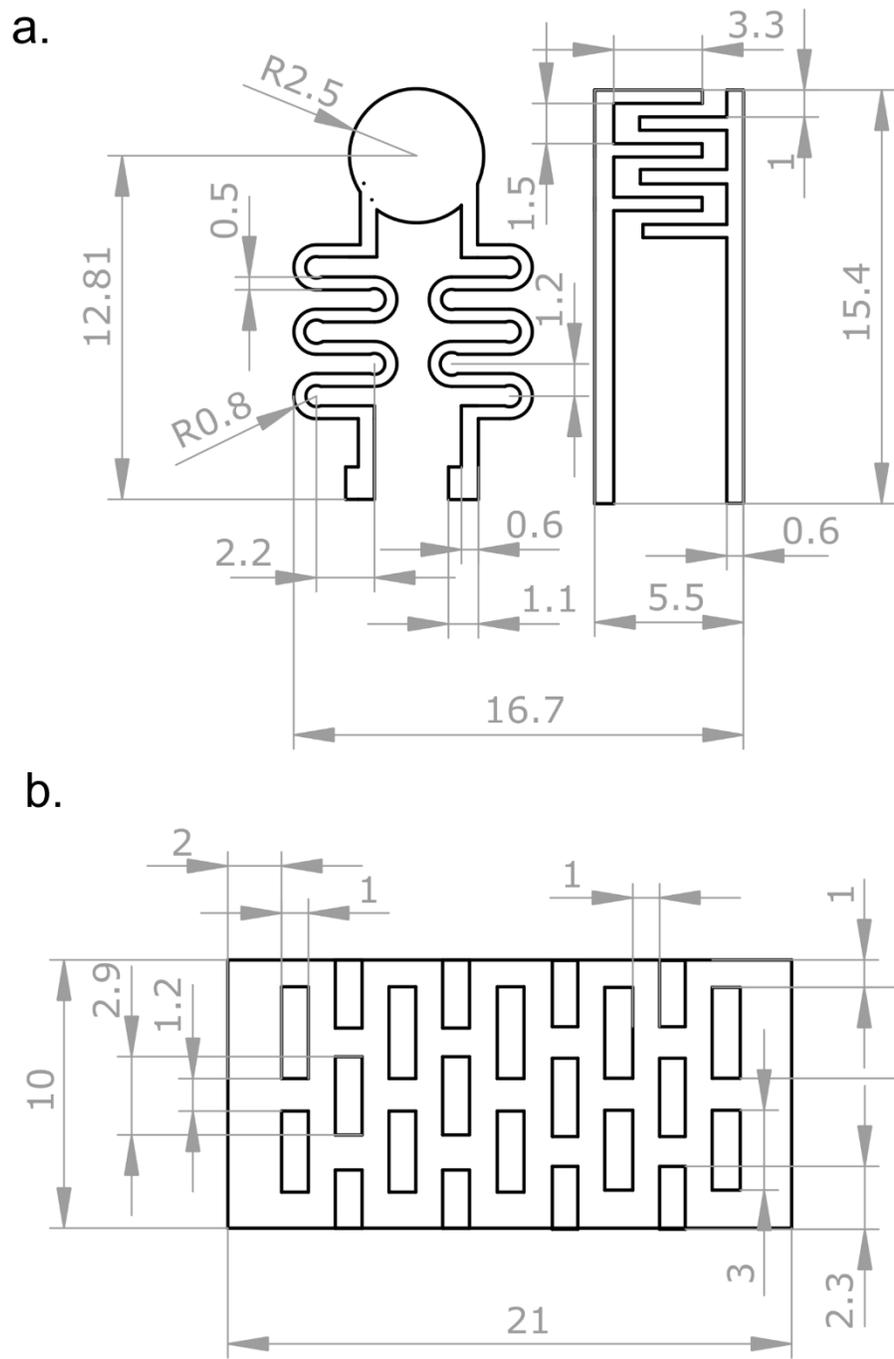

**Figure S3:** Dimension of the Leaf Tattoo (a) and Stem strain sensor exhibiting Kirigami structure (b). Note: All dimensions are in millimeters.

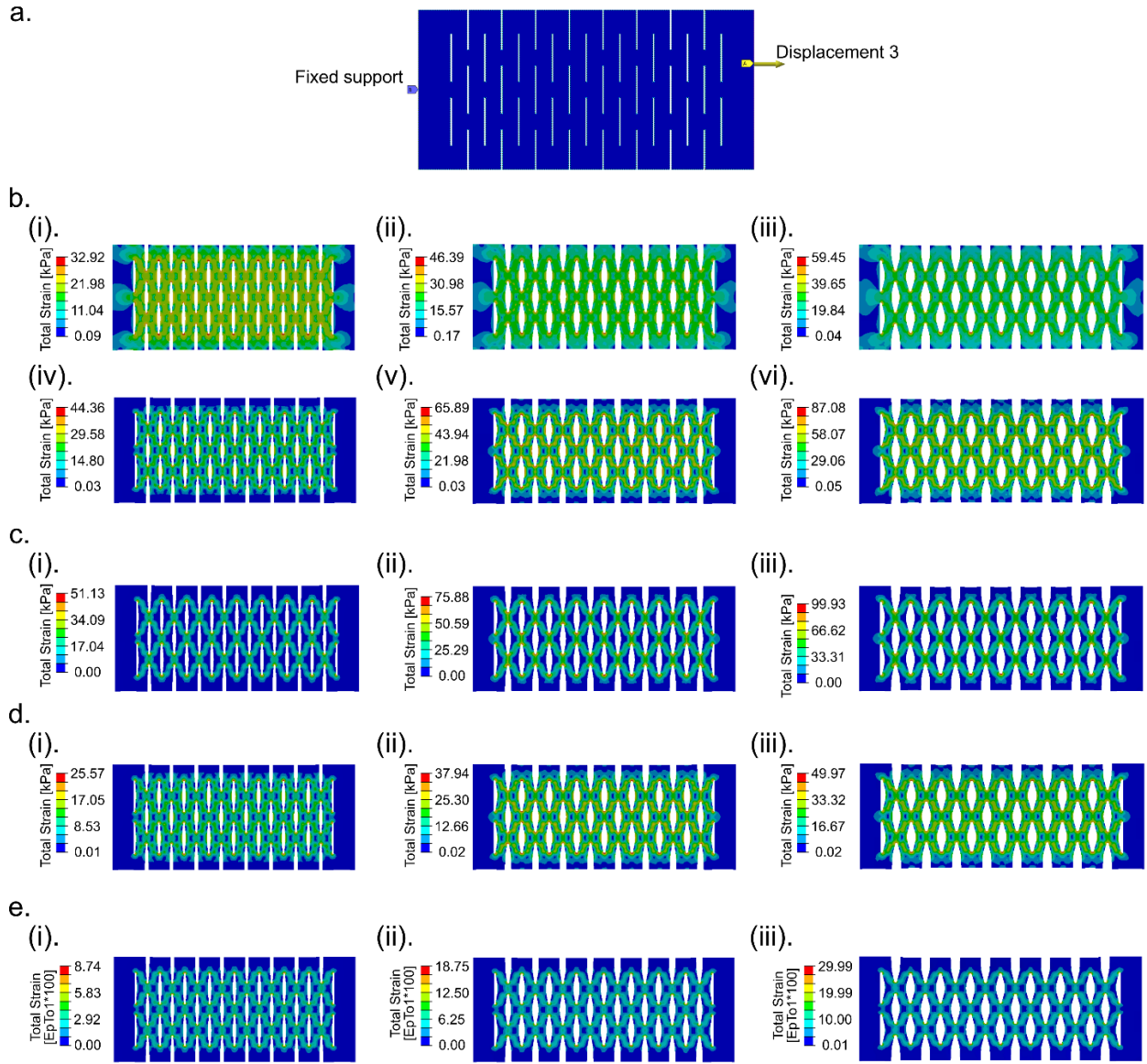

**Figure S4:** Kirigami-based strain sensor in Stem Tattoo numerical simulation using Ansys Software. (a) Boundary conditions applied on the Kirigami structure; (b) Maximum and minimum equivalent stress of polyimide at 1.0s (i), 2.0s (ii), & 3.0 s(iii) and eutectic gel film (on top of the polyimide sheet) at (iv) 1.0s (v) 2.0s, & (v) 3.0s ; (c) maximum principle stress of Kirigami strain sensor at (i) 1.0s, (ii) 2.0s, & (iii) 3.0 s; (d) maximum shear stress of Kirigami strain sensor at (i) 1.0 s, (ii) 2.0s & (iii) 3.0s; (e) equivalent strain at (i) 1.0s, (ii) 2.0 s, & (iii) 3.0s.

**Note S1: FEM-Based Analysis of the Leaf Tattoo and Eutectogel Kirigami Strain Sensor**

The FEM analysis of the leaf tattoo is conducted for bending and tension loading using two separate static structural configurations in Ansys 2022.R1 software. The model consisted of two layers for each material: a 0.1 mm thick tattoo carrier paper layer and a 0.05 mm thick graphene layer, both of which are considered bonded. The material properties of the tattoo carrier paper use values in reported literature [38,39].

A 0.25 mm element size mesh was applied using the MultiZone Quad/Tri Method for both configurations. "Capture Curvature" and "Capture Proximity" were enabled with default settings, generating identical meshes for the leaf tattoo and Eutectogel Kirigami strain sensor. In the bending setup, the left edge was fixed, and a remote fixed point at the right edge received a gradual 4.2 cm displacement along the negative x-axis over 1.75 s. For tension, a 0.2 cm displacement along the positive x-axis was applied gradually over 2.00 s.

The maximum and minimum equivalent strain at 0.5s for bending is measured to be $1.8959 \times 10^{-2}$ and $5.5278 \times 10^{-6}$; and at 0.5s, for tension is estimated to be $1.7708 \times 10^{-1}$ and $3.4799 \times 10^{-4}$ (Figure 2c (i) and figure 2c (v)). At 1.0s, these values were modified to be for bending $3.9612 \times 10^{-2}$ (max.) and $4.1713 \times 10^{-5}$ (min.); where in the case of the tension, it is found to be $3.7924 \times 10^{-1}$ (max.) and $4.0214 \times 10^{-4}$ (min.) as shown in Figure 2c(ii) and figure 2c(vi). Similarly, at 1.5s, the values become for bending $6.6283 \times 10^{-2}$ (max.) and $6.4866 \times 10^{-5}$ (min.); where for the tension, it is found to be $5.7574 \times 10^{-1}$ (max.) and $9.0293 \times 10^{-4}$ (min.) as illustrated in Figure 2c(iii) and figure 2c(vii). Finally, as seen in Figure 2c(iv) and Figure 2c(viii), at the end of time, for bending, the value of equivalent strain becomes $8.2792 \times 10^{-2}$ (max.) and $7.6073 \times 10^{-5}$ (min.), while the tension is found to be $7.6019 \times 10^{-1}$ (max.) and $1.0459 \times 10^{-3}$ (min.). These results suggest that the serpentine design effectively mitigated large strains when significant bending and tension were applied to the tattoo.

Finite Element Method (FEM) analysis for the eutectogel Kirigami-based strain sensor for stem sensors was conducted, considering both static structural and plastic deformation. The model included two layers for each material: a 125 μm thick PI layer and an 80 μm thick eutectic gel film layer, both of which were considered bonded. The dimensions of the Kirigami structure are shown in Figure S2b. Materials were modeled as non-linear, incorporating isotropic elasticity and bilinear isotropic hardening properties. The material properties of PI and eutectic gel were obtained from previous reports. [37,42].

Two boundary conditions are chosen for the model: (1) fixed support on the left edge and (2) displacement on the right edge. A displacement of 6 mm in the x-axis was configured to apply gradually over the 3s time (2 mm/s speed), as shown in Figure S4a. The maximum and minimum equivalent stress for polyimide is measured to be 32.918 kPa and 0.09461 kPa at 1.0 s, 46.385 kPa and 0.17415 kPa at 2.0 s, 59.45 kPa and 0.03942 kPa at 3.0s; and for eutectic gel film the values found to be 44.361 kPa and 0.025517 kPa at 1s, 65.893 kPa and 0.029075 kPa at 2s, 87.082 kPa and 0.045552 kPa at 3s (Figure S4b). On the other hand, the maximum and minimum principal stress for eutectic gel film are 51.13 kPa and $-4.4162 \times 10^{-7}$ kPa at 1s, 75.881 kPa, and $-3.8834 \times 10^{-4}$ kPa at 2s, 99.93 kPa, and $-1.2221 \times 10^{-3}$ kPa (Figure S4c). Similarly, the maximum and minimum shear stress for eutectic gel film is 25.568 kPa and $1.2806 \times 10^{-2}$ kPa at 1s, 37.939 kPa and $1.6339 \times 10^{-2}$ kPa at 2s, 49.965 kPa and $2.22975 \times 10^{-2}$ kPa (Figure S4d). The maximum and minimum equivalent strains at 1.0s is found to be 8.739 and 0.0033272, while at 2.0s, it is 18.752 and 0.0044132, and at 3s, it is 29.994 and 0.0060131 (Figure S4e). These results indicate that the Kirigami-patterned eutectogel structure is suitable for strain sensing.

a.

b.

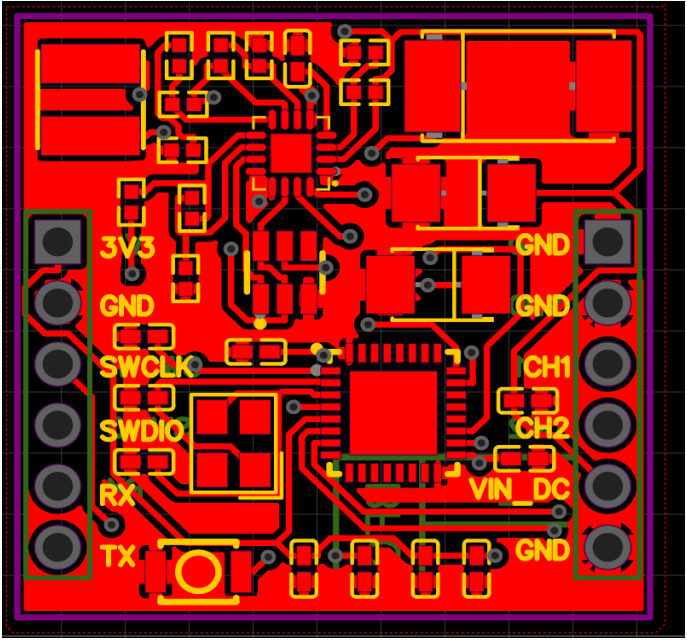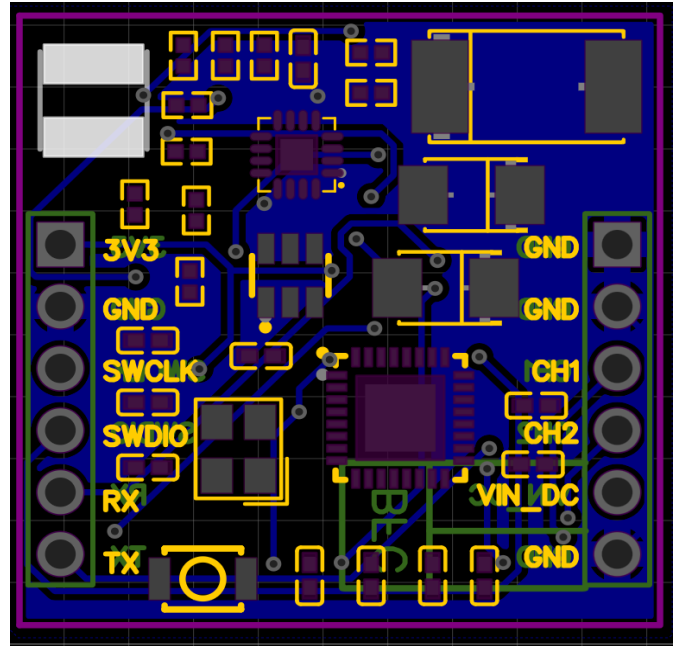

**Figure S5**: PCB layout of the development board. (a) Top Layer; (b) Bottom Layer

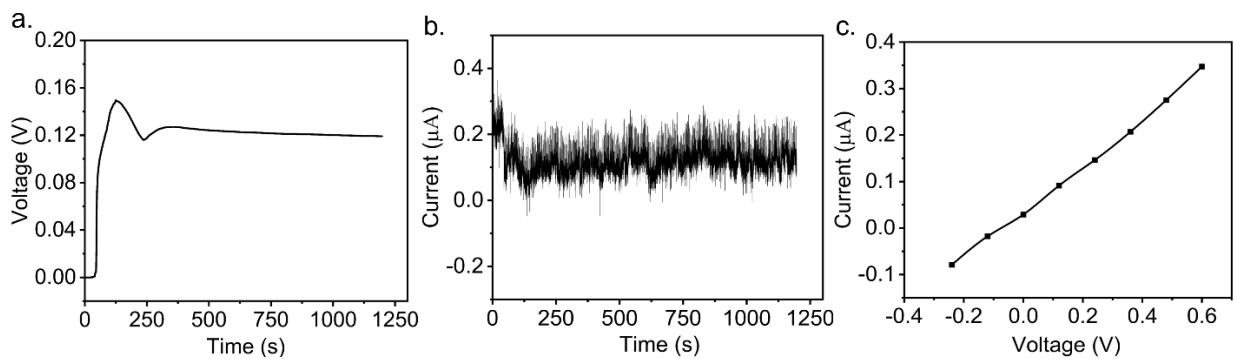

**Figure S6:** I-V characteristics of the $V_2O_5$ MEG; (a) Open circuit voltage, (b) short circuit current, and (c) IV characteristics of the $V_2O_5$-based MEG

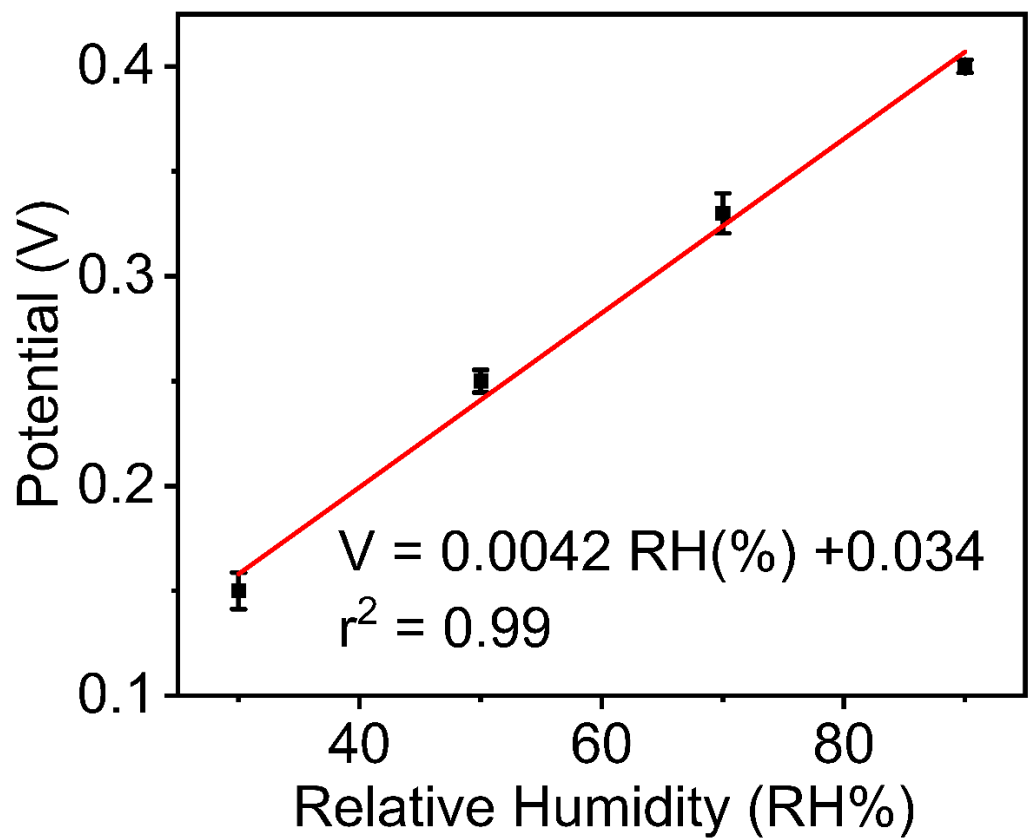

**Figure S7:** Humidity sensor calibration curve generated from MEG output voltage

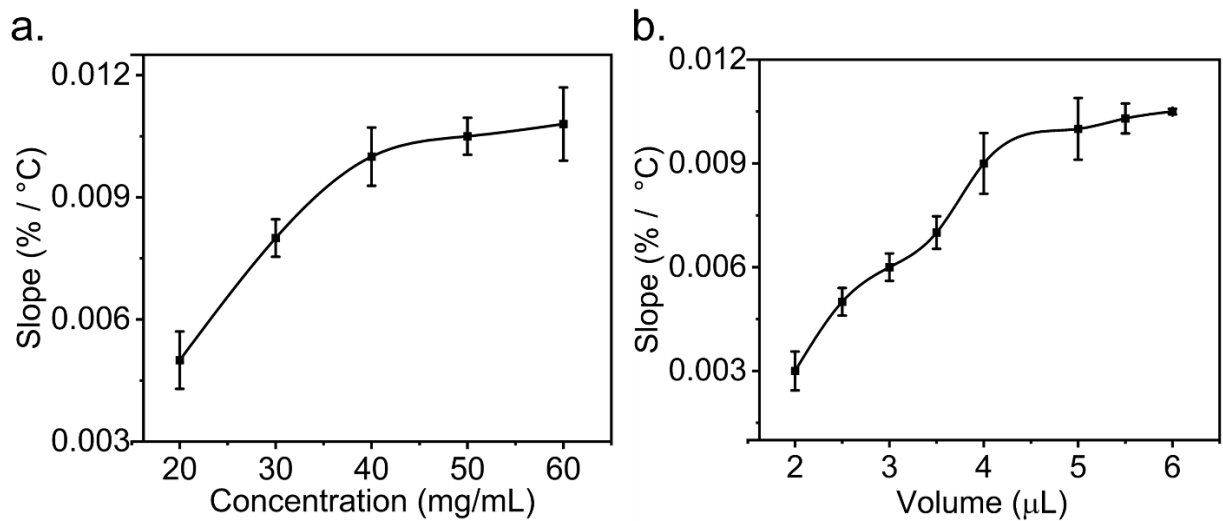

**Figure S8:** Temperature sensor Optimization; (a) concentration optimization (with constant volume of 5.0 µL); (b) volume optimization (with concentration of 50.0 mg/mL)

a. 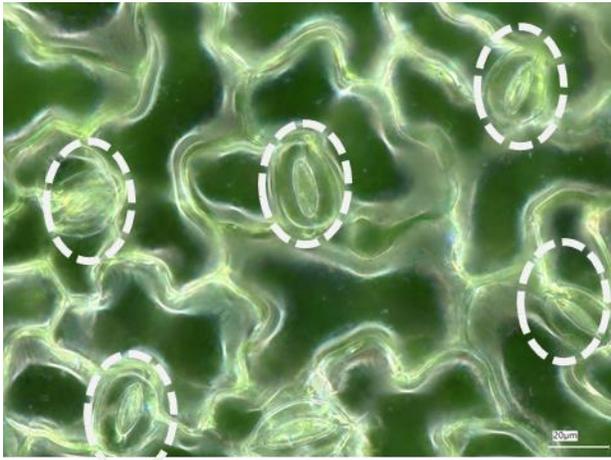 b. 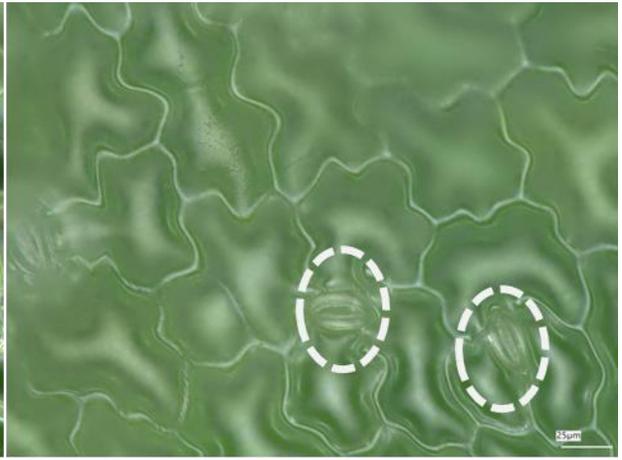

**Figure S9:** (a) Opening and (b) Closure of the pepper leaf stomata. The opening image of the stomata is from the control plant (healthy). The closed stomata are from water-stressed plants.

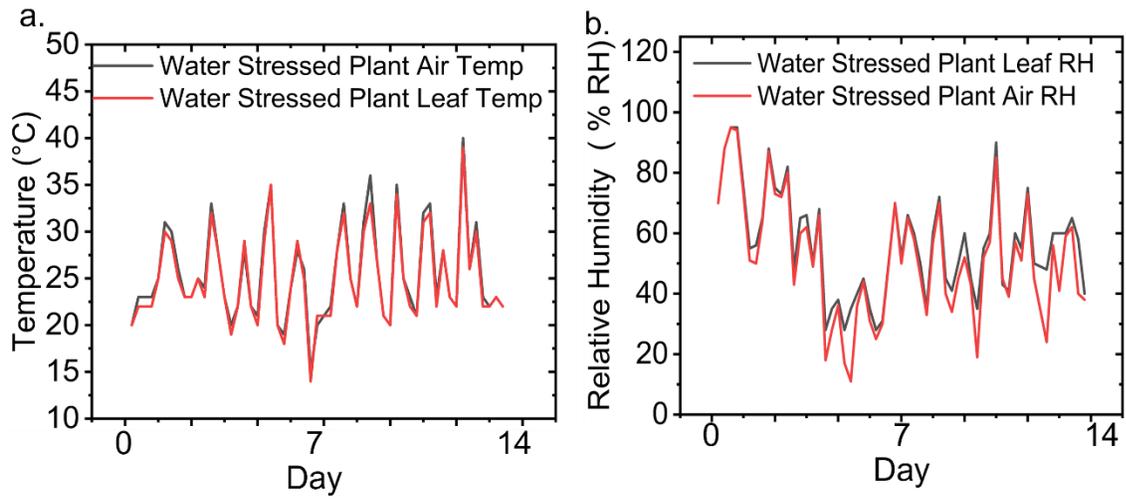

**Figure S10:** (a) Real-time water-stressed plant air and beneath-leaf temperature; (b) Real-time water-stressed plant beneath a leaf and open-air relative humidity

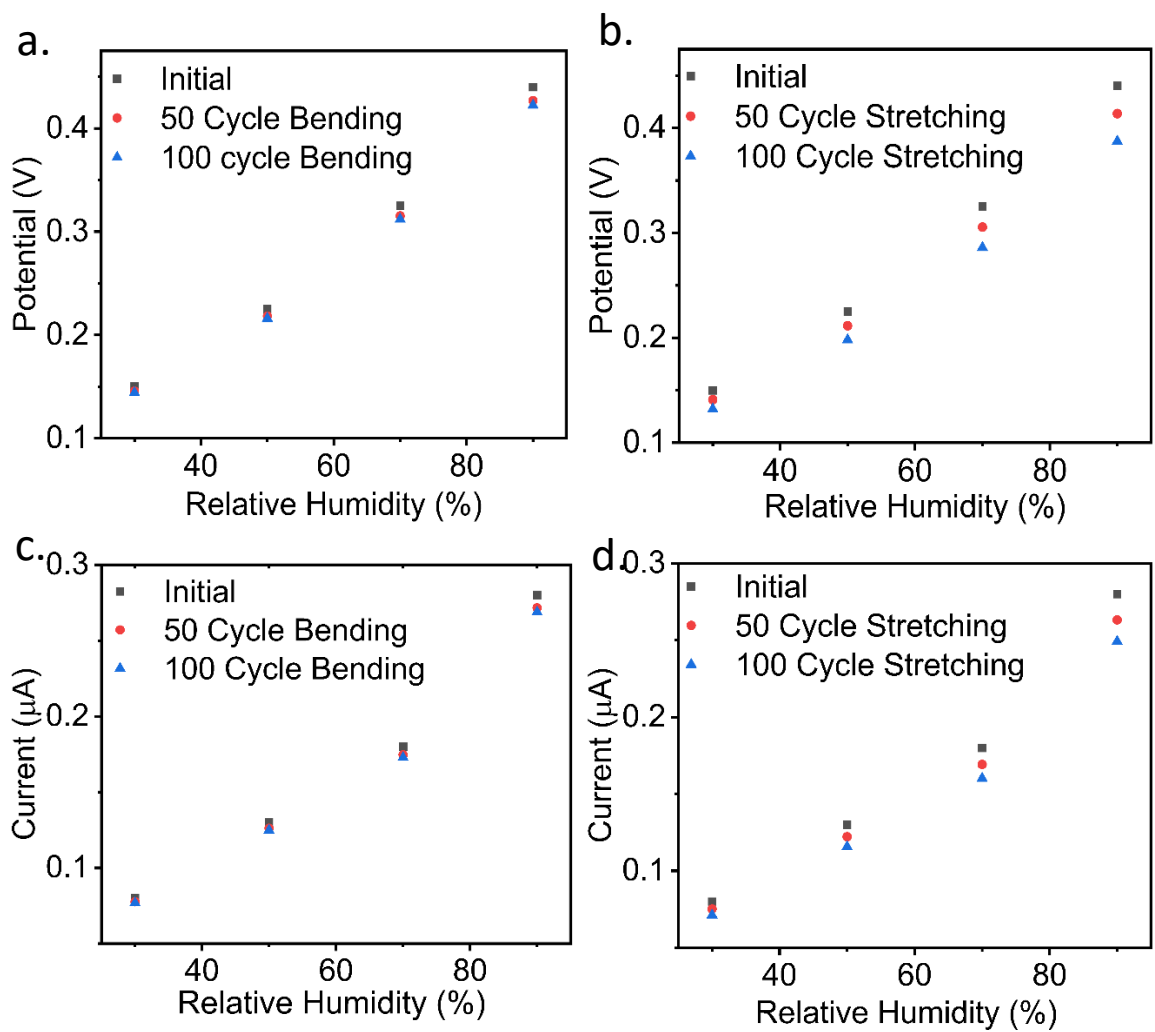

**Figure S11:** Bending (a) and stretching (b) response of the tattoo MEG's voltage generation. Bending (c) and stretching (d) response of the MEG's current generation.

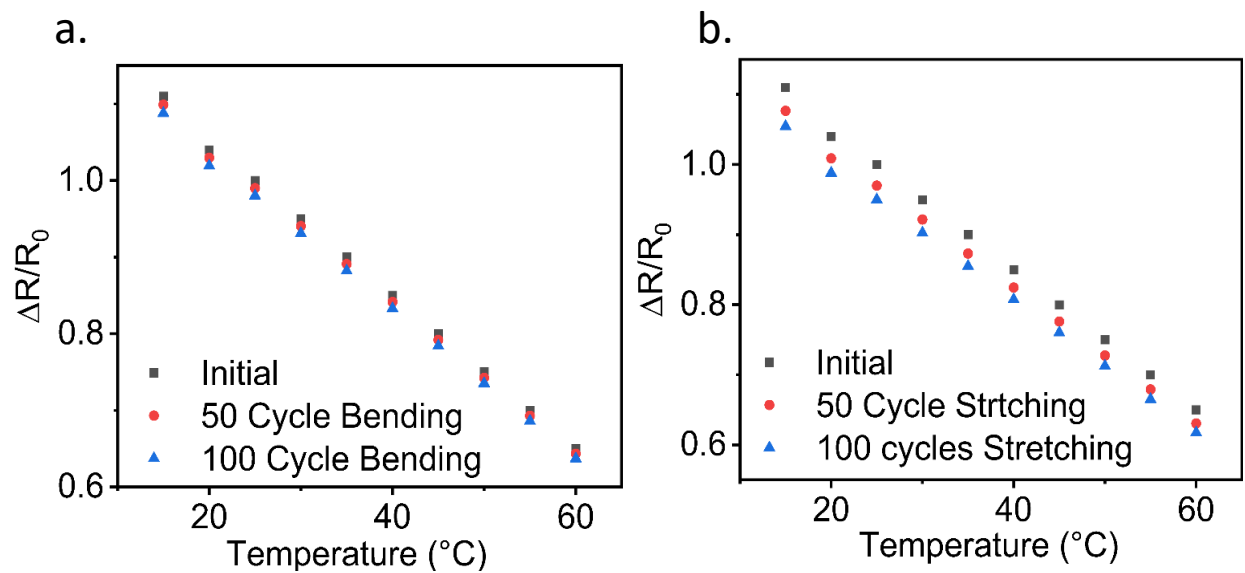

**Figure S12**. Bending (a) and stretching (b) of the tattoo Temperature Sensor.

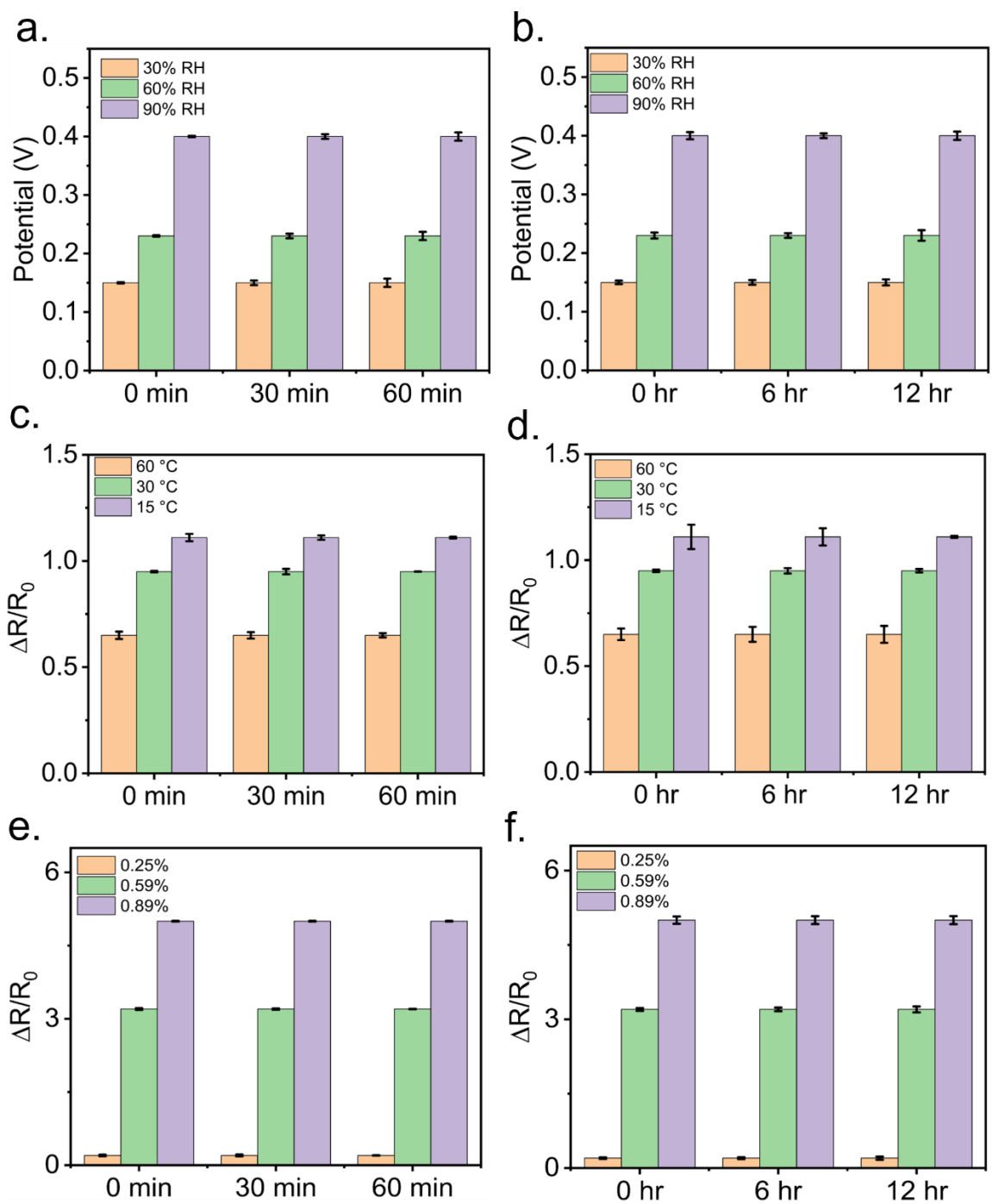

**Figure S13**: Stability Analysis of the Sensor. (a) Short-term (60 min) and (b) long-term (12 hr) stability of the MEG-based humidity sensor. (c) Short-term (60 min) and (d) long-term (12 hr) stability of the Temperature sensor. (e) Short-term (60 min) and (f) long-term (12 hr) stability of the MEG-based Kirigami-based strain sensor (here bending strain is shown).

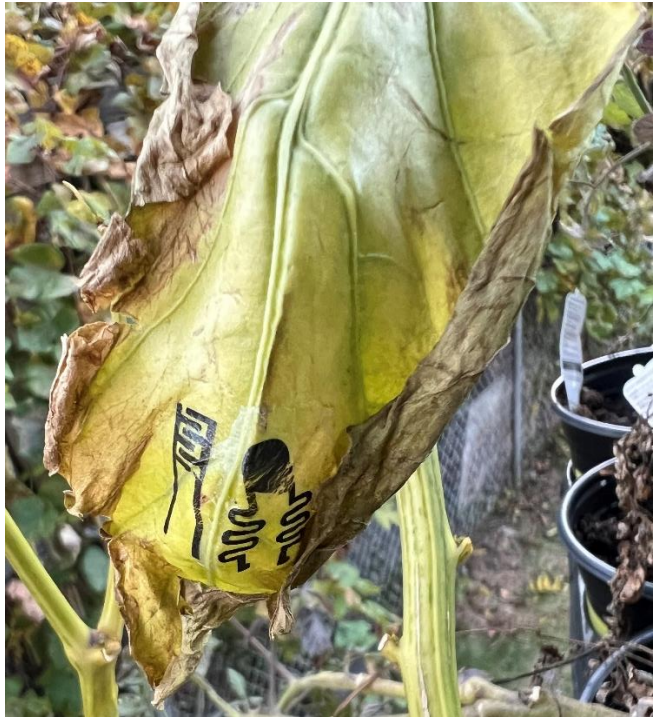

**Figure S14:** After 45 days, the plant leaf and the tattoo condition.

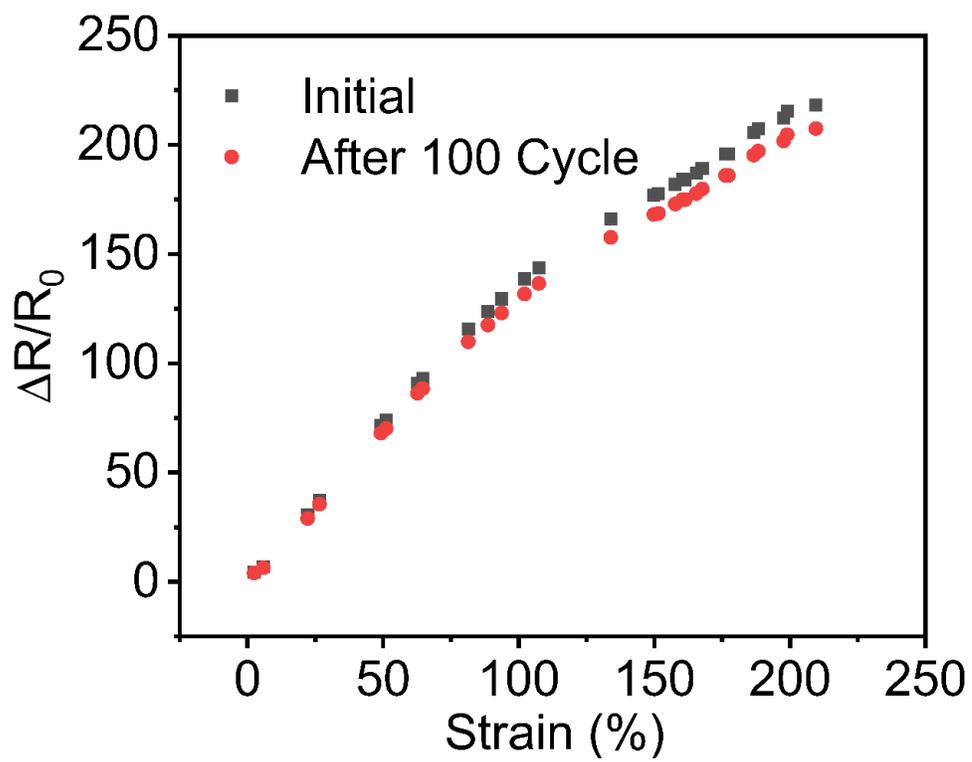

**Figure S15:** Kirigami Strain Sensor's response after 100 Cycles of Starching.

**Supplementary Table S1**: Real-time stem diameter data for the unstressed, water-stressed, and salinity-stressed plants.

| Day | Unstressed plant diameter in mm | | Water-stressed plant diameter in mm | | Salinity-stressed plant diameter in mm | |
|---|---|---|---|---|---|---|
| | Pristine Sensor | 100-cycle stretched sensor | Pristine Sensor | 100-cycle stretched sensor | Pristine Sensor | 100-cycle stretched sensor |
| 1 | 6.43 | 6.42 | 6.87 | 6.855 | 6.52 | 6.503 |
| 2 | 6.43 | 6.42 | 6.87 | 6.855 | 6.52 | 6.503 |
| 3 | 6.44 | 6.43 | 6.87 | 6.855 | 6.52 | 6.503 |
| 4 | 6.45 | 6.44 | 6.85 | 6.835 | 6.5 | 6.483 |
| 5 | 6.45 | 6.44 | 6.84 | 6.825 | 6.49 | 6.473 |
| 6 | 6.46 | 6.45 | 6.83 | 6.815 | 6.49 | 6.473 |
| 7 | 6.47 | 6.46 | 6.81 | 6.795 | 6.47 | 6.453 |
| 8 | 6.5 | 6.49 | 6.8 | 6.785 | 6.46 | 6.443 |
| 9 | 6.53 | 6.52 | 6.79 | 6.775 | 6.45 | 6.433 |
| 10 | 6.54 | 6.53 | 6.78 | 6.765 | 6.44 | 6.423 |
| 11 | 6.57 | 6.56 | 6.75 | 6.735 | 6.42 | 6.403 |
| 12 | 6.58 | 6.57 | 6.75 | 6.735 | 6.42 | 6.403 |
| 13 | 6.61 | 6.6 | 6.75 | 6.735 | 6.42 | 6.403 |
| 14 | 6.61 | 6.6 | 6.74 | 6.725 | 6.41 | 6.393 |
| 15 | 6.61 | 6.6 | 6.73 | 6.715 | 6.4 | 6.383 |
| 16 | 6.61 | 6.6 | 6.72 | 6.705 | 6.39 | 6.373 |
| 17 | 6.62 | 6.61 | 6.71 | 6.695 | 6.38 | 6.363 |
| 18 | 6.62 | 6.61 | 6.7 | 6.685 | 6.37 | 6.353 |
| 19 | 6.63 | 6.62 | 6.7 | 6.685 | 6.37 | 6.353 |
| 20 | 6.63 | 6.62 | 6.7 | 6.685 | 6.37 | 6.353 |
| 21 | 6.63 | 6.62 | 6.69 | 6.675 | 6.36 | 6.343 |
| 22 | 6.64 | 6.63 | 6.69 | 6.675 | 6.36 | 6.343 |

| 23 | 6.65 | 6.64 | 6.68 | 6.665 | 6.35 | 6.333 |
|---|---|---|---|---|---|---|
| 24 | 6.66 | 6.65 | 6.68 | 6.665 | 6.35 | 6.333 |
| 25 | 6.67 | 6.66 | 6.68 | 6.665 | 6.35 | 6.333 |
| 26 | 6.68 | 6.67 | 6.68 | 6.665 | 6.35 | 6.333 |
| 27 | 6.68 | 6.67 | 6.65 | 6.635 | 6.32 | 6.303 |
| 28 | 6.68 | 6.67 | 6.64 | 6.625 | 6.31 | 6.293 |
| 29 | 6.71 | 6.7 | 6.61 | 6.595 | 6.29 | 6.273 |
| 30 | 6.72 | 6.71 | 6.6 | 6.585 | 6.28 | 6.263 |
| 31 | 6.73 | 6.72 | 6.57 | 6.555 | 6.25 | 6.233 |
| 32 | 6.74 | 6.73 | 6.54 | 6.525 | 6.22 | 6.203 |
| 33 | 6.76 | 6.75 | 6.53 | 6.515 | 6.21 | 6.193 |
| 34 | 6.77 | 6.76 | 6.52 | 6.505 | 6.2 | 6.183 |
| 35 | 6.78 | 6.77 | 6.52 | 6.505 | 6.2 | 6.183 |
| 36 | 6.8 | 6.79 | 6.5 | 6.485 | 6.18 | 6.163 |
| 37 | 6.8 | 6.79 | 6.48 | 6.465 | 6.16 | 6.143 |
| 38 | 6.8 | 6.79 | 6.48 | 6.465 | 6.16 | 6.143 |
| 39 | 6.8 | 6.79 | 6.48 | 6.465 | 6.16 | 6.143 |
| 40 | 6.8 | 6.79 | 6.47 | 6.455 | 6.15 | 6.133 |